%% file: main.tex
\documentclass[%
 reprint,
 amsmath,amssymb,
 aps,
]{revtex4-2}

\usepackage{graphicx}
\usepackage{stackengine}
\usepackage{dcolumn}
\usepackage{bm}
\usepackage{lipsum}
\usepackage{natbib}
\usepackage{booktabs}
\usepackage{amsmath} 
\usepackage{textcomp} 
\usepackage{url}  
\usepackage[utf8]{inputenc}
\usepackage[english]{babel}
\usepackage{caption}
\usepackage{subcaption}
\usepackage{float}
\usepackage{mathtools}
\usepackage{tabularx}

\usepackage{SIunits}
\usepackage{graphicx}
\usepackage{xspace}
\usepackage{braket}
\usepackage{placeins}
\usepackage{leftidx}
\usepackage[makeroom]{cancel}

\usepackage[utf8]{inputenc}

\usepackage[usenames,dvipsnames]{color}

\newcommand{\SP}{\gamma}

\usepackage[pdftex, colorlinks=true, urlcolor=blue, linkcolor=blue, citecolor=blue, pdfstartview=FitH, plainpages=false]{hyperref}

\usepackage[utf8]{inputenc}

\begin{document}

\input{macros.tex}

\preprint{APS/123-QED}

\title{First-principles calculations of transport coefficients in Weyl semimetal TaAs}

\author{Guillaume E. Allemand}
\email{guillaume.allemand@uliege.be}
 \affiliation{Nanomat group, Q-MAT, University of Liège,  and European Theoretical Spectroscopy Facility}
 
\author{Matteo Giantomassi}
\affiliation{Institute of Condensed Matter and Nanosciences, 
Université\ Catholique\ de\ Louvain,\ Louvain-la-Neuve, Belgium}
\affiliation{European Theoretical Spectroscopic Facility (ETSF)}
 
\author{Matthieu J. Verstraete}%
 
\affiliation{Nanomat group, Q-MAT, University of Liège,  and European Theoretical Spectroscopy Facility}%
\affiliation{ITP, Physics Department, Utrecht University 3508 TA Utrecht, The Netherlands}




\date{\today}

\begin{abstract}
We study charge and heat transport from first-principles in the topological Weyl semimetal TaAs. Electron-phonon coupling matrix elements are calculated using density functional perturbation theory and used to derive the thermo-electric transport coefficients,  including the electrical conductivity, Seebeck coefficient, electronic thermal conductivity and the Peltier coefficient.
We compare the self-energy and momentum relaxation time approximations to the iterative solution of the Boltzmann Transport Equation, finding they give similar results for TaAs provided the chemical potential is treated accurately. 
For the iterative method, we derive an additional equation, which is needed to fully solve for transport under both thermal and an electrical potential gradients. Interestingly, the Onsager reciprocity between $S$ and $\Pi$ is no longer imposed, and we can deal with systems breaking time-reversal symmetry, in particular magnetic materials.
We compare our results with the available experimental data  for TaAs: the agreement is excellent for $\sigma_{xx}$, while $\sigma_{zz}$ is overestimated, probably due to differences in experimental carrier concentrations. The Seebeck coefficient is of the same order of magnitude in theory and experiments, and we find that its low-T behavior also strongly depends on the doping level.
\end{abstract}


                              
\maketitle

\section{\label{sec:level1}Introduction}
\subsection{Context and motivations}

Transport in topologically non-trivial materials is an important development of recent condensed matter physics, and one of the main methods able to probe the fundamental and delicate nature of these systems. 
Weyl semimetals (WSM) are crystalline topological materials in which the electrons near the Fermi level behave as massless chiral fermions (known as Weyl fermions).
The complex electronic structure of WSM restricts their phase space, and leads to the protection of certain carrier states from scattering, and, as a consequence, high mobility. 
Specific signatures of topology can appear in the thermoelectric response, such as the chiral anomaly in WSM, which results in a strong positive magneto-conductance that can be detected experimentally~\cite{huang2015observation}.
Beyond fundamental characterization of topology, the exceptional and/or exotic transport properties of WSM are of great interest for industrial applications, and more specifically for thermoelectric functionality. The latter is now widely used in electronic and spintronic devices, in autonomous sensors and for waste heat scavenging. 

%
%
 
In this work, we focus on the prototypical WSM, tantalum arsenide, the first experimentally proven Weyl semimetal~\cite{lv2015experimental}.
A few previous works~\cite{huang2015observation,xiang2017anisotropic, xu2021thermoelectric} have characterized the (magneto) thermoelectric properties of TaAs experimentally. 
Several theoretical studies of the electron and electron-phonon interactions in TaAs also exist in the literature, in particular Garcia~\cite{garcia2020optoelectronic}, Coulter~\cite{coulter2019uncovering}, and Peng~\cite{peng2016high}.
Garcia and Coulter characterize the electron phonon coupling in detail and the optoelectronic response, but not transport.
Peng et al. calculate the transport coefficients, but they solve the Boltzmann Transport Equation (BTE) using the constant relaxation time approximation (cRTA), which is known to fail in metals~\cite{xu2020thermoelectric}, and cannot capture the subtleties of low-energy scattering around the Fermi level.

In this study, we characterize the thermoelectric transport properties of TaAs in a fully first-principles way, to account for chemical bonding, detailed band dispersion and scattering effects. We determine the equilibrium geometry and ground-state electronic structure, and compute the electron-phonon coupling (EPC) matrix elements in order to obtain the transport coefficients by solving the BTE. 
We use the~\textsc{ABINIT} software suite~\cite{Gonze2020}, performing Density Functional Theory (DFT)~\cite{Hohenberg1964,Kohn1965}, Density Functional Perturbation Theory (DFPT)~\cite{Gonze1997, Baroni2001}, EPC and transport calculations.
Our results for the conductivity in plane closely match the experimental values, while the results for the Seebeck coefficients overestimate the measurements of Ref.~\cite{xiang2017anisotropic}. Both show a strong dependency on the doping present in the samples. 
Additionally, we present our calculated electronic thermal conductivities, which are challenging to compare with the experimental values, as the latter are \emph{estimated} 
indirectly. 
We also discuss the reciprocity of the off-diagonal Onsager coefficients in the BTE by comparing the Seebeck and Peltier coefficients.
Finally, we show that the Wiedemann-Franz relation can be broken strongly for Weyl semimetals such as TaAs, in a range of temperature from around 150 K to 350 K.

\subsection{\label{sec:level2}Structure and computational methods}

TaAs crystallizes in a 4-atom primitive cell with a tetragonal structure (I41md, space group \#109), and non-magnetic electronic configuration. 
The calculations are performed with fully-relativistic norm-conserving GGA-PBE~\cite{PBE1} pseudopotentials~\cite{Hamann2013,VANSETTEN201839} including spin-orbit coupling, an energy cutoff of 50 Ha, a 16x16x16 $\Gamma$-centered $\kk$-point grid, and a 8x8x8 $\qq$-point grid for phonons.
Our lattice constants ($a = 3.465, c=11.743$ \AA) are in good agreement with previous theoretical results ($a=3.467, c= 11.755$ \AA), but slightly overestimate experimental values ($a=3.437, c= 11.656$ \AA)~\cite{chang2016phonon}, as expected when using GGA.
%

\section{Electron and phonon band structures}

%

\begin{figure}[thb]
            \centering
            \captionsetup{type=figure}
            \stackinset{r}{0.6cm}{b}{0.7cm}{\includegraphics[height=2.5cm]{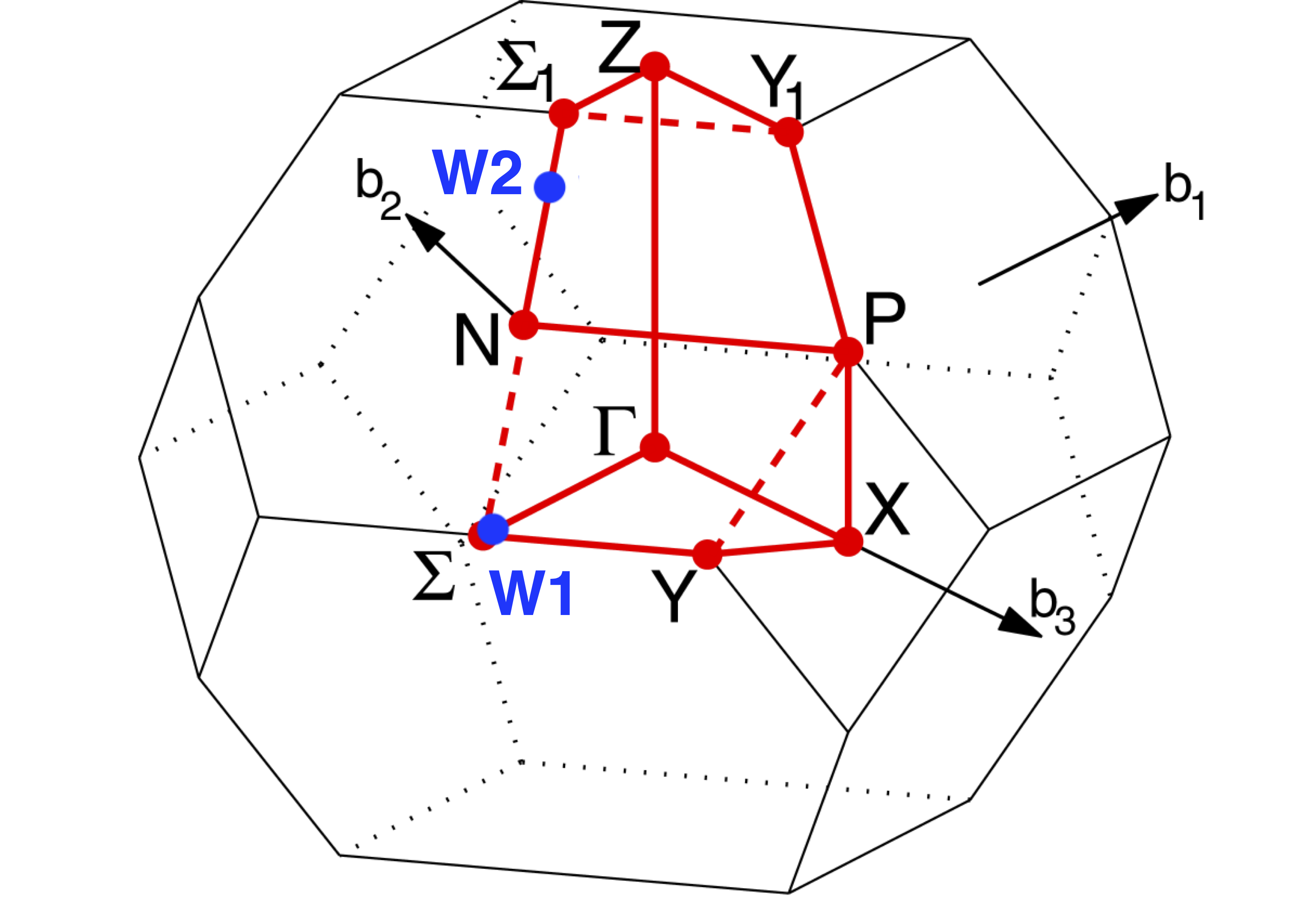}}{\includegraphics[scale=0.68]{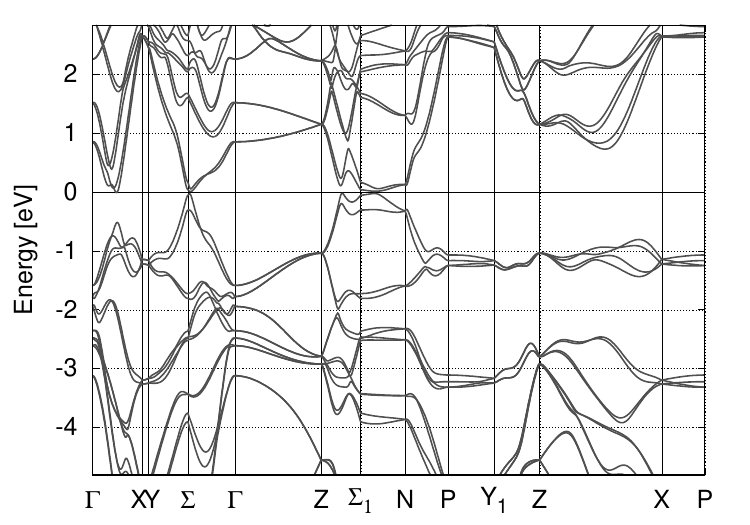}}
           \caption{\label{figure:elec}Electronic band structure of TaAs within GGA-PBE, taking into account spin-orbit coupling. The two types of Weyl points are visible near $\Sigma$ and $\Sigma_1$. Inset adapted from Ref.~\cite{SETYAWAN2010299}.}
\end{figure}

The electronic band structure calculated along a high-symmetry $\kk$-path is shown in Figure~\ref{figure:elec}. The valence and conduction bands cross in the vicinity of the Fermi level (set to 0 eV) at band crossings known as Weyl nodes. Two inequivalent nodes are contained in the band structure and TaAs exhibits 24 Weyl nodes in total. 
The shape of the band structure is characteristic of Weyl semimetals: direct band gaps throughout the Brillouin Zone, except at specific points dubbed W1 and W2. Our band dispersion is in good agreement with previous first-principles calculations~\cite{lee2015fermi}.

%

\begin{figure}[thb]
            \centering
            \captionsetup{type=figure}
            \includegraphics[scale=0.7]{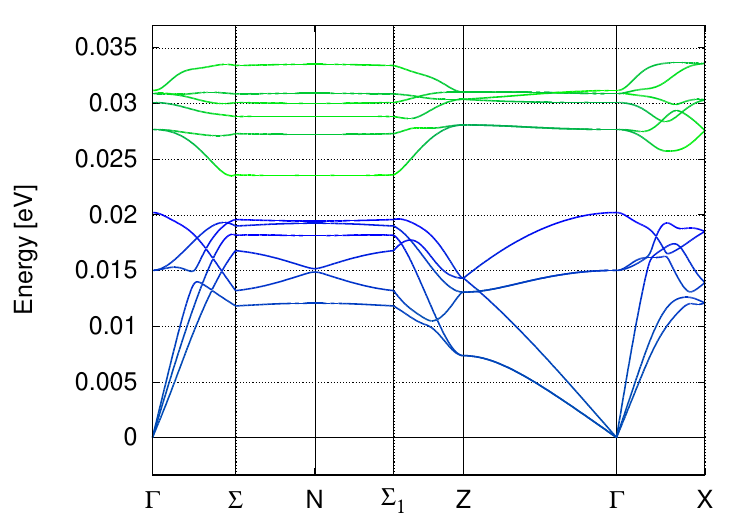}
            \caption{\label{figure:phonons}Phonon dispersion curve of TaAs. The green color corresponds to the contribution of As atoms and blue to the contribution of heavier Ta atoms.}
\end{figure}

The phonon dispersion is shown in Figure~\ref{figure:phonons}. 
There is no gap between acoustic and optical modes, but there are two main manifolds. Indeed, heavier Ta atoms contribute mainly at low frequencies (in blue) and As atoms at high frequencies (in green). 
Our results agree well with experimental and theoretical data reported in~\cite{chang2016phonon}. Our phonon frequencies at $\Gamma$ are
0.00; 14.6; 20.2; 27.6; 29.9; 30.0; 31.2 meV.

\section{Thermoelectric transport calculations}

In order to obtain the transport coefficients of TaAs, we compute the EPC matrix elements and solve the linearized Boltzmann Transport Equation. The results presented in this article are obtained either using  Relaxation Time Approximation (RTA) solutions to the BTE (Self-Energy RTA or Momentum RTA), or using the Iterative Boltzmann Transport Equation (IBTE)~\cite{Giustino2017}. The methods implemented within the EPH module of \textsc{ABINIT} allow us to calculate the basic electronic transport coefficients: the electrical conductivity $\sigma$ or resistivity ($\rho$), the Seebeck coefficient ($S$), the electronic thermal conductivity ($\kel$) and the Peltier coefficient (${\Pi}$). 
For a detailed description of the EPH code and RTA implementation, see Ref.~\cite{Brunin2020b, Claes2022} and for a review of the different approaches to solving the BTE, see Ref.~\cite{ponce2020first}. In this section, we summarize the SERTA/MRTA methods, and present an extension of the IBTE formalism that we have implemented in a development version of \textsc{ABINIT} 10.3.
Atomic Hartree units are used throughout the paper. We note, in particular, that linewidths (inverse lifetimes) require a $1/\hbar$ prefactor, which is simply 1 in atomic units.

\subsection{Boltzmann transport equation}

The Boltzmann equation is a semi-classical theory in which the system is assumed to have well-defined quasiparticle excitations (with crystalline momentum $\kk$, band index $n$ and group velocity $\vv_\nk$) of energy $\ee_\nk$ with negligible imaginary part,
meaning the spectral function in the Bloch band picture is strongly peaked around the quasiparticle energy.
The fundamental aim of the theory is the determination of the out-of-equilibrium statistical distribution function $f_n(\rr,\kk)$, usually when a steady state is reached.
The \emph{time-independent} Boltzmann transport equation (BTE) for the steady state is expressed as:
\begin{equation}\label{eq:initial_BTE_steady}
-\PDER{f_n}{\rr} \cdot \vv_\nk 
-\PDER{f_n}{\kk} \cdot \bmcF
+  \mcI_{n\kk}[f] = 0
\end{equation}
where the first term is diffusive and enters into play if the external fields have a spatial dependence, while the second term is the so-called drift contribution, under a generic force $\bmcF$. 
In Eq.~\ref{eq:initial_BTE_steady}, 
$\mcI_{n\kk}[f]$ is the scattering integral, a functional of $f$ that represents the net number of particles entering/leaving the infinitesimal phase space region around $(\rr, \kk)$ due to scattering processes. Once the out-of-equilibrium distribution function $f$ is known, one can compute the electric current density:
\begin{equation}
\label{eq:def_je}
\jj^e = -\dfrac{\SP e}{\Omega N_\kk} \sum_{n\kk} \vv_\nk f_\nk
\end{equation}
as well as the heat current density $\jj^Q$:
\begin{equation}
\label{eq:def_jQ}
\jj^Q = \dfrac{\SP}{\Omega N_\kk} \sum_{n\kk} (\ee_\nk -\mu) \vv_\nk f_\nk 
\end{equation}
where the factor $\SP$ accounts for spin degeneracy ($\SP = 2$ in spin unpolarized systems and $\SP = 1$ for spin polarized systems or spinor wavefunctions when spin-orbit coupling is taken into account), 
$\Omega$ is the volume of the unit cell, 
$N_\kk$ the number of $\kk$-points in the Brillouin zone (BZ), $\mu$ is the chemical potential,
and $e = |e|$ the absolute value of the electronic charge (+1 in atomic units).

The complexity of Eq.~\ref{eq:initial_BTE_steady} stems from the mathematical structure of the scattering functional. To overcome this difficulty, one often uses the BTE for a steady state in the linear regime, for a weak electric field $\EE$ and thermal gradient $\TT$, and eventually an additional magnetic field $\BB$, producing $\bmcF = -e (\EE + \vv_\nk \times \BB)$. Since the external fields are weak,
one can assume that the solution of the BTE can be expressed as a small correction to the equilibrium Fermi-Dirac occupation function according to:
\begin{equation}\label{eq:fermi}
f_\nk = f^0(\ee_\nk, \mu, T) + \delta f_\nk(T, \mu)
\end{equation}
where the unknown $\delta f_\nk$ term is linear in the external fields and $f^0(\ee_\nk, \mu, T)$ is the local equilibrium distribution. 
Note that only the $\delta f_\nk$ terms contribute 
to $\jj^e$ and $\jj^Q$
since $f^0$ is an even function of $\kk$ while $\vv_\nk$ is an odd function.
The scattering integral is then linearized as follows:
\begin{equation}
\mcI_\nk[f] \approx
\mcI_\nk[f^0] + \Lnk = 
\Lnk 
\end{equation} 
where we used $\mcI_\nk[f^0] = 0$, since $f^0$ is a solution of the BTE at zero external fields.
After some algebra, one obtains the \emph{linearized} version of the BTE:
\begin{equation}\label{eq:linearizedBTE}
\begin{multlined}
\dfodenk \vv_\nk \cdot
\biggl [
-e \biggl(\EE + \dfrac{1}{e}\PDER{\mu}{\rr}\biggr) - \dfrac{(\ee_\nk-\mu)}{T}  \PDER{T}{\rr} 
\biggr ] \\ + 
\vv_\nk \cdot \PDER{\delta f_\nk}{\rr} +
\bmcF^B_\nk \cdot \PDER{\delta f_\nk}{\kk}
=  \mcL_{n\kk}[f^0, \delta f]
\end{multlined}
\end{equation}
where $\bmcF^B_\nk$ is the Lorentz force $-e \vv_\nk \times \BB$.
This linearized expression
represents the starting point of both the RTA and IBTE formalisms that are discussed in more detail in the next sections.

At this point, it is important to highlight that, in the framework of linear response theory within non-equilibrium thermodynamics, the steady-state current densities and the driving fields are related to each other via Onsager's relations (written with our own notations, see~\cite{behnia2015fundamentals} for more details):
\begin{equation}
\label{eq:onsager}
\begin{pmatrix}
\jj^e\\\jj^Q
\end{pmatrix}
=
\begin{pmatrix}
\LL_{11}(\BB) & \LL_{12}(\BB) \\
\LL_{21}(\BB) & \LL_{22}(\BB)
\end{pmatrix}
\cdot
\begin{pmatrix}
\bmcE \\
\dfrac{-\nabla T}{T}
\end{pmatrix}
\end{equation}
where $\bmcE = \EE + \nabla_\rr \mu / e$ is the electrochemical potential (we only consider the $\EE$ contribution in our calculations below), the matrix elements $\LL_{ij}(\BB)$ are called Onsager coefficients and are transport tensors that depend on the external magnetic field $\BB$. These tensors inherit the symmetry properties of the underlying crystal lattice, meaning certain components may vanish or become equivalent due to symmetry constraints, significantly simplifying the analysis in high-symmetry systems.
The explicit dependence on $\BB$ in Eq.~\ref{eq:onsager} emphasizes that an external magnetic field can modify both electrical and thermal conductivities. 
According to Onsager’s reciprocity principle, which arises from the time-reversal symmetry of microscopic dynamics, the transport coefficients satisfy the following relationship:
\begin{equation}
\LL_{12}(\BB) = \LL_{21}(-\BB).
\end{equation}
In the absence of time reversal symmetry breaking (\(\BB = 0\)), 
this reduces to the symmetric form:
\begin{equation}
\label{eq:reci_onager}
\LL_{12} = \LL_{21}.
\end{equation}
In the following sections, we will review the various methodologies commonly employed to solve the linearized BTE and assess the extent to which Onsager’s reciprocal relation is preserved in each approach.

\subsection{Relaxation-time approximation}

Within the relaxation-time approximation (RTA), the scattering integral $\mcL_{n\kk}$ in Eq.~\ref{eq:linearizedBTE} is replaced by the rather simple expression
\begin{equation}
\Lnk =  -\dfrac{\delta f_\nk}{\tau_\nk}.
\end{equation} 
From a physical perspective, we are assuming that 
once the external fields are switched off, the 
$f_\nk$ occupation function relaxes to the equilibrium $f^0_\nk$ following an exponential decay with a (single, averaged) time constant $\tau_\nk$, and the $f_\nk$ relaxes as though all other occupations were at equilibrium. This presumes that the occupations of different states $\nk$ do not influence each other, though this is never strictly correct.
From a mathematical perspective, the RTA drastically simplifies the initial problem, since we have replaced the linearized scattering integral with a much simpler expression that depends only on the relaxation time $\tau_\nk$.
The linearized BTE within the RTA (Eq.~\ref{eq:linearizedBTE}) is expressed as (we assume $\BB = 0$, and spatially constant $\EE$ and $\TT$ such that the  spatial gradient of $\delta f$ is 0):
\begin{equation}
\delta f_\nk = 
-\tau_\nk \dfodenk \vv_\nk \cdot
\biggl [
-e \bmcE - \dfrac{(\ee_\nk-\mu)}{T}  \PDER{T}{\rr}
\biggr ].
\end{equation}
%
 
%
By using this result, one can express the current densities (Eq.~\ref{eq:def_jQ} and Eq.~\ref{eq:def_je}) in terms of the external fields and finally
derive the expression of the Onsager transport tensors, using Eq.~\ref{eq:onsager}:
%
%
\begin{equation}
\LL_{11} = -\SP e^2 \sum_n \int \dkopi \dfodenk \tau_\nk \vv_\nk \otimes \vv_\nk
\end{equation}
\begin{equation}
\LL_{12} = \LL_{21} = \SP e \sum_n \int \dkopi \dfodenk \tau_\nk (\ee_\nk - \mu) \vv_\nk \otimes \vv_\nk
\end{equation}
\begin{equation}
\LL_{22} = -\SP \sum_n \int \dkopi \dfodenk \tau_\nk (\ee_\nk - \mu)^2 \vv_\nk \otimes \vv_\nk.
\end{equation}
%
The integral is performed over the $\kk$-points in the BZ region $\Omega_{\mathrm{BZ}}$ and $\vnk$ is the velocity operator computed with DFPT:
\begin{equation}
\vnk = \PDER{\enk}{\kk} = \langle \nk | \dfrac{\partial{H}}{\partial{\kk}} | \nk \rangle.
\end{equation}
%
In should be noted that, in the RTA, the Onsager reciprocal relation 
$\LL_{12}$ = $\LL_{21}$ is automatically fulfilled.

In the cRTA, $\tau_\nk$ is assumed to be a constant,
independent of crystalline momentum and band index.
This is clearly a rather crude approximation, especially for metals. In a more physically accurate approach, the relaxation time should depend on
the microscopic state of the electron and the temperature.
In the so-called Self-Energy Relaxation Time Approximation (SERTA), one employs Fermi's golden rule to compute 
the scattering rate due to the interaction with phonons up to first order in the displacement. 
This leads to the following expression:
\begin{equation}
\label{eq:serta}
\begin{split}
    \frac{1}{\tau_{n\kk}} & =
                2 \pi
                \sum_{m,\nu} \int_\BZ \frac{\dd\qq}{\Omega_{\mathrm{BZ}}} |\gkq|^2 
                \\
                & \times \left[ (n_\qnu + f_{m\kk+\qq})
                                \delta(\enk - \emkq  + \wqnu) \right.\\
                & \left. + (n_\qnu + 1 - f_{m\kk+\qq})
                                \delta(\enk - \emkq  - \wqnu ) \right],
\end{split}
\end{equation}
where $\gkq$ is the electron-phonon coupling matrix element, $n_\qnu$ the phonon distribution and $\wqnu$ the phonon frequency.
The use of the SERTA acronym is related to the fact that
Eq.~\ref{eq:serta} corresponds to twice the value of the imaginary part of the Fan-Migdal self-energy evaluated at the Kohn-Sham eigenvalue~\cite{Giustino2017}.
This result is derived in a more rigorous way in Appendix~\ref{appendix:scattering_operator}.
See also~\cite{ponce2020first, Brunin2020b, Giustino2017} for further details.
%
%
The SERTA ignores most back-scattering processes. 
In the MRTA, these are partially accounted for by expressing the transport lifetime as:
\begin{equation}
\label{eq:mrta}
\begin{split}
    \frac{1}{\tau_{n\kk}} & =
                2 \pi
                \sum_{m,\nu} \int_\BZ \frac{\dd\qq}{\Omega_{\mathrm{BZ}}} |\gkq|^2 \left( 1 - \frac{\vnk \cdot \vmkq}{|\vnk|^2} \right) \\
                & \times \left[ (n_\qnu + f_{m\kk+\qq})
                                \delta(\enk - \emkq  + \wqnu) \right.\\
                & \left. + (n_\qnu + 1 - f_{m\kk+\qq})
                                \delta(\enk - \emkq  - \wqnu ) \right].
\end{split}
\end{equation}

\subsection{Iterative BTE}

In this section, we review the iterative BTE solution, which is more accurate and goes beyond RTA by explicitly including the in and out e-ph scattering in the formalism (for more details, see~\cite{li2015electrical,ponce2020first,Claes2022}). 

Let us start by rewriting the linearized BTE (Eq.~\ref{eq:linearizedBTE}) for $\BB = 0$, assuming a constant (weak) electric field, a constant temperature gradient, and restricting ourselves to first-order derivatives:
\begin{equation}
\label{eq:ibte_full}
\dfodenk \vv_\nk 
\biggl [
-e \bmcE  -\dfrac{(\ee_\nk - \mu)}{T} \PDER{T}{\rr}
\biggr ]  = -\dfrac{\delta f_\nk}{\tau^0_\nk} + \Leph.
\end{equation}
The right-hand side (compared to Eq.~\ref{eq:linearizedBTE}) has a first term equivalent to the RTA formalism, with $\tau^0_\nk$ given by the Fan-Migdal diagram, plus a second term explicitly taking into account the linearized e-ph scattering integral. The derivation of this scattering integral can be found in~\cite{ponce2020first} and is summarized with our notation in the appendix.

We also express the first-order correction to the distribution function (Eq.~\ref{eq:fermi}) as a first perturbation linear in $\EE$ ($\delta f^\bmcE_\nk$) and a second linear in $\TT$ $(\delta f^T_\nk$):
\begin{equation}
\label{eq:deltaF_seebeck}
\delta f_\nk 
= \delta f^\bmcE_\nk + \delta f^T_\nk = 
\FF^\bmcE_\nk \cdot \EE 
+ \FF^T_\nk \cdot \TT / T %
\end{equation}
where $\FF^T_\nk$ and $\FF^\bmcE_\nk$ are tensors giving the first derivatives of the distribution function $f_\nk$ with respect to $\TT$ and $\EE$, respectively (as derived in the relaxation time context by Fiorentini and Bonini~\cite{fiorentini2016}). $\FF^T_\nk$ is obtained by solving the IBTE at zero $\EE$-field, whereas $\FF^\bmcE_\nk$ is obtained at zero temperature gradient ($\TT = 0$).
The IBTE equations are then solved using an iterative scheme, and read:
%
\begin{equation}\label{eq:FE}
\begin{multlined}
\FF^{\bmcE \ {(i+1)}}_\nk = e \dfodenk \tau^0_\nk \vv_\nk \\
+ \dfrac{\tau^0_\nk 2 \pi}{N_\qq}
\sum_{m\qnu} |\gkq|^2 \\
\times \bigl [
(1 + n_\qnu - f^0_{n\kk}) \delta^+ + 
(n_\qnu + f^0_{n\kk}) \delta^- 
\bigr ]
\FF^{\bmcE \ {(i)}}_{m\kq}
\end{multlined}
\end{equation}
%
and
\begin{equation}\label{eq:FT}
\begin{multlined}
\FF^{T \ {(i+1)}}_\nk = 
\dfodenk \tau^0_\nk \vv_\nk (\ee_\nk - \mu) \\
+ \dfrac{\tau^0_\nk 2 \pi}{N_\qq}
\sum_{m\qnu} |\gkq|^2 \\
\times \bigl [
(1 + n_\qnu - f^0_{n\kk}) \delta^+ + 
(n_\qnu + f^0_{n\kk}) \delta^- 
\bigr ]
\FF^{T \ {(i)}}_{m\kq}
\end{multlined}
\end{equation}
%
where $N_\qq$ is the number of $\qq$-points in the BZ and $\delta^\pm = \delta(\enk - \emkq \pm \wqnu)$. 
Eq.~\ref{eq:FE} is solved with the initial guess $\FF^{\bmcE \ {(0)}}_\nk = \tau^0_\nk \vv_\nk$ while Eq.~\ref{eq:FT} is solved with the initial guess $\FF^{T \ {(0)}}_\nk = \frac{(\ee_\nk -\mu)}{e} \FF^\bmcE_\nk $. The latter is a full solution for the redundant $\FF^{T}$ in the RTA case which does not have the second term in the RHS of Eq.~\ref{eq:ibte_full} . The Ansatz for $\FF^{T}$ is not exact in the full IBTE, as there is a mismatch of the bands $m$ and $n$ in the scattering operator.

\subsection{Transport coefficients within IBTE}
\label{subsect:transp_coeff}

In order to calculate the transport coefficients in the IBTE formalism, we make use of the Onsager relations (Eq.~\ref{eq:onsager}):
\begin{eqnarray}
\label{eq:je_IBTE}
\jj^{e} &=& \LL_{11} \cdot \EE - \LL_{12} \cdot \frac{\TT}{T} \\
\jj^{Q} &=& \LL_{21} \cdot \EE - \LL_{22} \cdot \frac{\TT}{T}= - \boldsymbol{\kel} \cdot \TT
\label{eq:jq_IBTE}
\end{eqnarray}
and we compare them to the current density expressions within the IBTE obtained by inserting Eq.~\ref{eq:deltaF_seebeck} in Eq.~\ref{eq:def_je} and in Eq.~\ref{eq:def_jQ}. As above, $f^0(\ee_\nk, \mu, T)$ from Eq.~\ref{eq:fermi} does not contribute to the density currents:
\begin{equation}
\jj^e \!\! = \!\! -\dfrac{\SP e}
{\Omega N_\kk} 
\sum_{n\kk} \vv_\nk \left(\FF^\bmcE_\nk \cdot \EE + \FF^T_\nk \cdot \dfrac{\TT}{T} \right)
\end{equation}
\begin{equation}
\jj^Q \!\! = \!\! \dfrac{\SP}
{\Omega N_\kk} 
\sum_{n\kk} (\ee_\nk \! -\! \mu) \vv_\nk\, \!\! \left(\FF^\bmcE_\nk \cdot \EE + \FF^T_\nk \cdot \dfrac{\TT}{T} \right).    
\end{equation}
Therefore, we can derive the expression of the Onsager coefficients within IBTE:
\begin{eqnarray}
\label{eq:Onsag_coeff}
\LL_{11} &=& - \dfrac{\SP e^2}{\Omega N_\kk} \sum_\nk \vv_\nk \otimes \FF^\bmcE_\nk \\
\label{eq:L12}
\LL_{12} &=& \dfrac{\SP e}{\Omega N_\kk} \sum_{n\kk} \vv_\nk \otimes \FF^T_\nk \\
\label{eq:L21}
\LL_{21} &=& \dfrac{\SP}{\Omega N_\kk} \sum_{n\kk} (\ee_\nk -\mu) \vv_\nk \otimes \FF^\bmcE_\nk \\
\LL_{22} &=& - \dfrac{\SP}{\Omega N_\kk} \sum_{n\kk} (\ee_\nk -\mu) \vv_\nk \otimes \FF^T_\nk \label{eq:Onsag_fin}
\end{eqnarray}
%
The outer product is needed to preserve the dependency of $\FF^\bmcE$ and $\FF^T$ on the direction of the applied fields ($\EE$ or $\TT$), so each $\LL_{ij}$ is a tensor of rank two. 
We note that now the Onsager reciprocity is not imposed. With the current form for the scattering matrix, the initial $F^{T}$ obtained from $F^\bmcE$ cannot be an exact solution.
For time-reversal symmetric cases (like TaAs below) the breaking of Onsager reciprocity tests the numerical convergence and IBTE resolution scheme. 
In cases without time-reversal symmetry, we can now evaluate both $S$ and $\Pi$ explicitly. 

Finally, by considering the definitions of the transport tensors (see~\cite{behnia2015fundamentals}) and by replacing the Onsager coefficients in Eq.~\ref{eq:je_IBTE} and Eq.~\ref{eq:jq_IBTE} with their expressions in Eq.~\ref{eq:Onsag_coeff} to~\ref{eq:Onsag_fin}, we are now able to express the electrical conductivity ($\boldsymbol{\sigma}$), the Seebeck coefficient ($\mathbf{S}$), the electrical resistivity ($\boldsymbol{\rho}$), the electronic thermal conductivity $\boldsymbol{\kel}$ and the Peltier coefficient ($\boldsymbol{\Pi}$) within the IBTE formalism.

\section{Results and discussions}

\subsection{Numerical parameters for EPC}

In order to calculate the transport coefficients, one needs the EPC matrix elements $\gkq$ on a fine grid of wave-vectors, both in $\kk$- and $\qq$-space.
To do so, we use the interpolation method for the first-order change of the KS potential, as described in~\cite{Eiguren2008, Brunin2020b, Gonze2020} that allows us to
reach $\qq$-meshes that are much denser than those used in the initial DFPT calculation.
For the description of the electronic states, we \emph{explicitly} compute the KS wavefunctions within an energy window around the Fermi level, as these are the states that contribute to the transport coefficients.
The procedure is described in~\cite{Brunin2020b, Claes2022} and is briefly summarized in what follows.

In the first step, we employ the star-function interpolation method by Shankland-Koelling-Wood~\cite{Shankland1971,Koelling1986,Madsen2006,Madsen2018}.
This method takes as input a set of eigenvalues from the irreducible wedge of the BZ and a single parameter defining the basis set, allowing us to 
predict whether wavevectors of a much denser $\kk$-mesh fall within the energy window without having to solve the KS eigenvalue problem exactly. 
Next, we compute the KS wavefunctions non-self-consistently
for the relevant $\kk$-points.
These wavefunctions are finally used to compute the EPC matrix elements and transport properties.
Specifically, we consider an energy window of $0.25$ eV around the Fermi level,  
an interpolated fine grid of 64$\times$64$\times$64 $\qq$-points for the scattering potentials
and the same (filtered) $\kk$-point grid for the wavefunctions. A tutorial for the Abinit software details the practical calculation workflow with restricted energy ranges.
The Fermi surface filtering retains only 3178 $\kk$-points and 4 bands in our case.
The conductivity values at room temperature differ by $5 \%$ along the x direction and $20 \%$ along the z axis, for two successive fine grids (56$\times$56$\times$56 and 64$\times$64$\times$64).

\subsection{Electrical conductivity} \label{subsection:sigma}

The electrical conductivity quantifies a material's ability to conduct electrical current. In the absence of a temperature gradient, considering Ohm's law $\jj^e = \sigma \EE$ microscopically and replacing Eq.~\ref{eq:Onsag_coeff} in Eq.~\ref{eq:je_IBTE}, one is able to express the electrical conductivity tensor as: 
\begin{equation}
\boldsymbol{\sigma} = \LL_{11} = - \dfrac{\SP e^2}{\Omega N_\kk} \sum_\nk \vv_\nk \otimes \FF^\bmcE_\nk
\label{eq:sigma_IBTE}
\end{equation}
and by definition, the resistivity tensor is simply $\boldsymbol{\rho} = \boldsymbol{\sigma}^{-1} $.

\begin{figure}[th]
     \centering
        \includegraphics[width=1\linewidth]{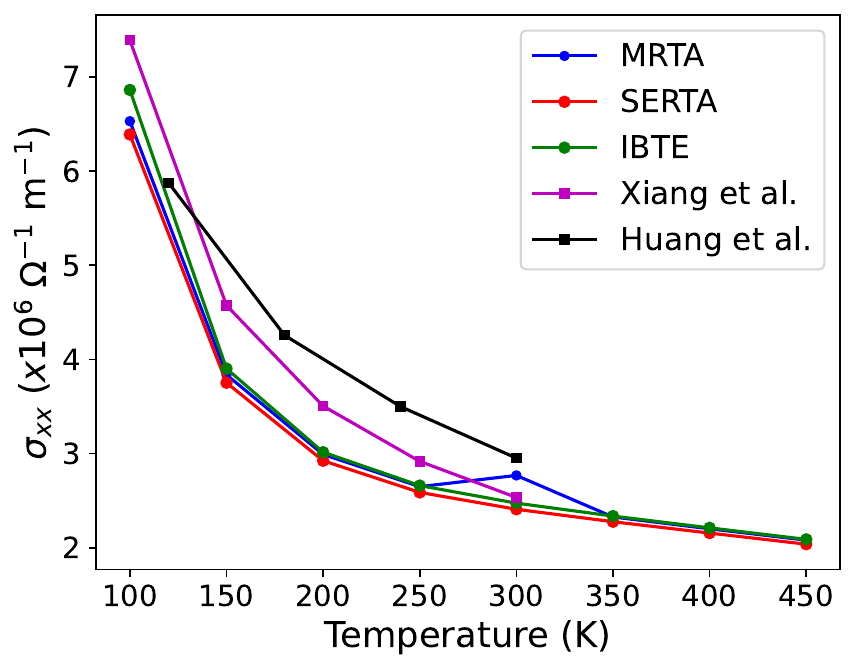}
        \includegraphics[width=1\linewidth]{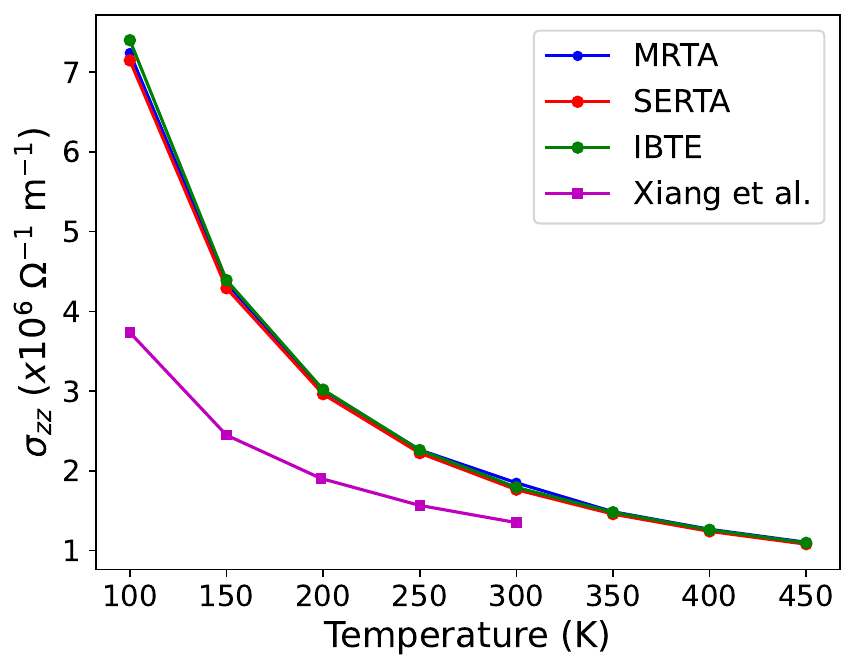}
        \caption{Temperature dependence of the xx and zz components of the electrical conductivity, calculated using RTA and IBTE methods. Experimental data come from~\cite{huang2015observation} and~\cite{xiang2017anisotropic}.}
        \label{fig:sigma}
\end{figure}

The calculated electrical conductivities for TaAs along the x and z directions are shown in Figure~\ref{fig:sigma}. We present conductivity values calculated following the three methods described above (SERTA, MRTA and IBTE), and find that, for TaAs, the different formalisms are quite close in both the x and z directions. This holds true for the other transport coefficients as well.

We compare to available experimental values reported by Xiang et al.~\cite{xiang2017anisotropic} and Huang et al.~\cite{huang2015observation}. Our results are in good agreement with both sets of experimental data for the conductivities along the x direction. Those along the z direction are larger than the experimental values, especially at low temperature. The explanation given in~\cite{xiang2017anisotropic} for their measured anisotropy in $\sigma$ is linked to the difference in effective mass and Fermi velocity between the x and z directions at the W2 point. These are automatically included in our calculations, but are not sufficient to produce such a strong anisotropy. 

The Hall coefficient measurements of Ref.~\cite{xiang2017anisotropic} also find a strongly anisotropic picture, with low-T Hall coefficient of 
$1.5$ $cm^{3}/C$ for current along the z axis and $-0.25$ $cm^{3}/C$ for the x axis. At room temperature, the z value converges to $0.15$ and the x Hall coefficient to $0.03$ $cm^{3}/C$. 
We note that the equivalence of $R_H$ in Ohm to a doping carrier concentration presumes the parabolic band approximation, which is particularly inappropriate in WSM.

To test this effect, we perform Hall coefficient calculations using the BoltzTrap2 code~\cite{BoltzTraP2}, and compare our results with Ref.~\cite{xiang2017anisotropic}. BoltzTrap2 includes the effects of electron band dispersion but employs the cRTA approximation with no lifetime variation between electron states.

For current along the x axis, we find $R_H$ values of 0.01 cm$^3$/C at 300 K, similar to the experiment. However, for current along the z axis, we obtain 0.02 cm$^3$/C, similar to x, but an order of magnitude smaller than Xiang et al.~\cite{xiang2017anisotropic}.
In our calculations, TaAs is quite isotropic both in $\sigma$ and $R_H$. The difference with experiment may result from the type and extent of the natural doping, or the details of the band structure within GGA DFT.
We discuss the consequences of doping within the rigid band approximation in the next subsection and in Appendix B.

\subsection{Seebeck coefficient}

The Seebeck coefficient relates a temperature difference to the resulting thermoelectric voltage, or, equivalently, the temperature gradient $\TT$ to the electric field $\EE$, measured in open circuit conditions ($\jj^e = 0$), through: $\mathbf{S} = \EE \otimes \frac{1}{\TT}$.
From Eq.~\ref{eq:je_IBTE}:
\begin{equation}
    \jj^{e} = \LL_{11} \cdot \EE - \LL_{12} \cdot \frac{\TT}{T} = 0
\end{equation}
which yields the Seebeck coefficient following its definition:
\begin{eqnarray}
     \mathbf{S} &=& \EE\otimes \frac{1}{\TT} = \frac{1}{T} (\LL_{11})^{-1} \LL_{12} \\
     &=&  \ \frac{\boldsymbol{\sigma}^{-1}}{T} \left( \dfrac{\SP e}{\Omega N_\kk} \sum_{n\kk} \vv_\nk \otimes \FF^T_\nk \right)
\end{eqnarray}
where we made use of Eq.~\ref{eq:sigma_IBTE} and Eq.~\ref{eq:L12}.
\begin{figure}[h]
        \includegraphics[width=1\linewidth]{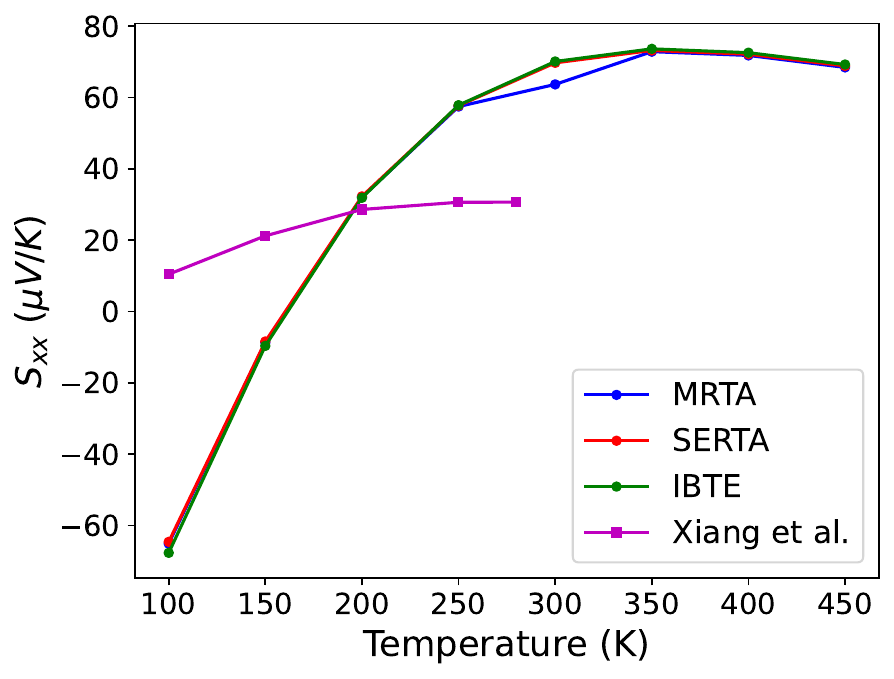}
        \includegraphics[width=1\linewidth]{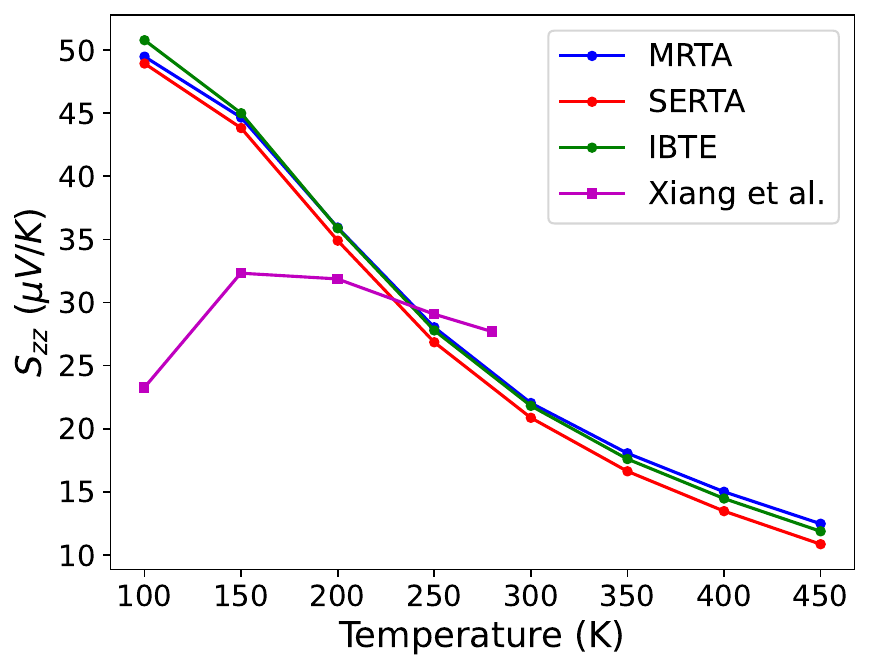}
        \caption{Temperature dependence of the xx and zz components of the Seebeck coefficient, calculated using RTA and IBTE methods. Experimental data come from~\cite{xiang2017anisotropic}.}
        \label{fig:SBK}
\end{figure}
The results for the Seebeck calculations are reported in Figure~\ref{fig:SBK}, again, along x and z directions. Our values are of the same order of magnitude as the experimental ones coming from~\cite{xiang2017anisotropic}, but the agreement is less quantitative than for $\sigma$, in particular for the x direction: our Seebeck coefficient becomes negative for temperatures below $\sim$200 K.
To test the relation between $S$ and carrier concentration, we performed calculations considering p-type doping (see Appendix B) and find that the value of $\mathbf{S_{xx}}$ becomes positive at low temperature, suggesting a strong effect of the natural doping.
As shown in~\cite{peng2016high}, $\mathbf{S}$ is indeed strongly dependent on the chemical potential value and is positive for p-type doping.

\subsection{Electronic thermal conductivity}

The thermal conductivity of a material is a measure of its ability to conduct heat through the transport of carriers or through phonons. By considering Eq.~\ref{eq:jq_IBTE}, imposing $\jj^{e}=0$, isolating $\EE$ in Eq.~\ref{eq:je_IBTE} and replacing it in the definition of $\boldsymbol{\kel}$, one obtains the carrier contribution to the thermal conductivity:
\begin{eqnarray}
- \boldsymbol{\kel} \ \TT &=& \LL_{21} (\LL_{11})^{-1} \LL_{12} \frac{\TT}{T} - \LL_{22} \frac{\TT}{T} \\
\boldsymbol{\kel} &=& \frac{\LL_{22}}{T} - \frac{1}{T} \LL_{21} (\LL_{11})^{-1} \LL_{12} \\
 &=& \frac{\LL_{22}}{T} - \LL_{21} \cdot \mathbf{S}.
\end{eqnarray}
Finally, by replacing $\LL_{21}$ and $\LL_{22}$ by their expressions (Eq.~\ref{eq:L21} and Eq.~\ref{eq:Onsag_fin}), the electronic thermal conductivity within the IBTE formalism is expressed as:
\begin{equation}
\begin{multlined}
\boldsymbol{\kel} = \dfrac{1}{T} \left( - \dfrac{\SP}{\Omega N_\kk} \sum_{n\kk} (\ee_\nk -\mu) \vv_\nk \otimes \FF^T_\nk \right) \\
- \left( \dfrac{\SP}{\Omega N_\kk} \sum_{n\kk} (\ee_\nk -\mu) \vv_\nk \otimes \FF^\bmcE_\nk \right) \cdot \mathbf{S}
\label{eq:kappa_IBTE}
\end{multlined}
\end{equation}
\begin{figure}[h]
     \centering
        \includegraphics[width=1\linewidth]{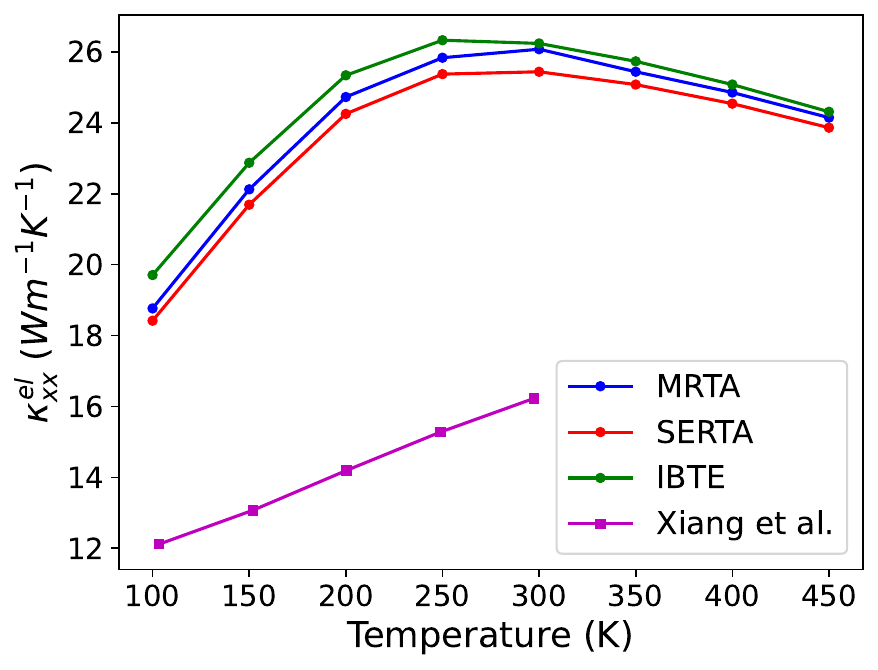}\hfill
        \includegraphics[width=1\linewidth]{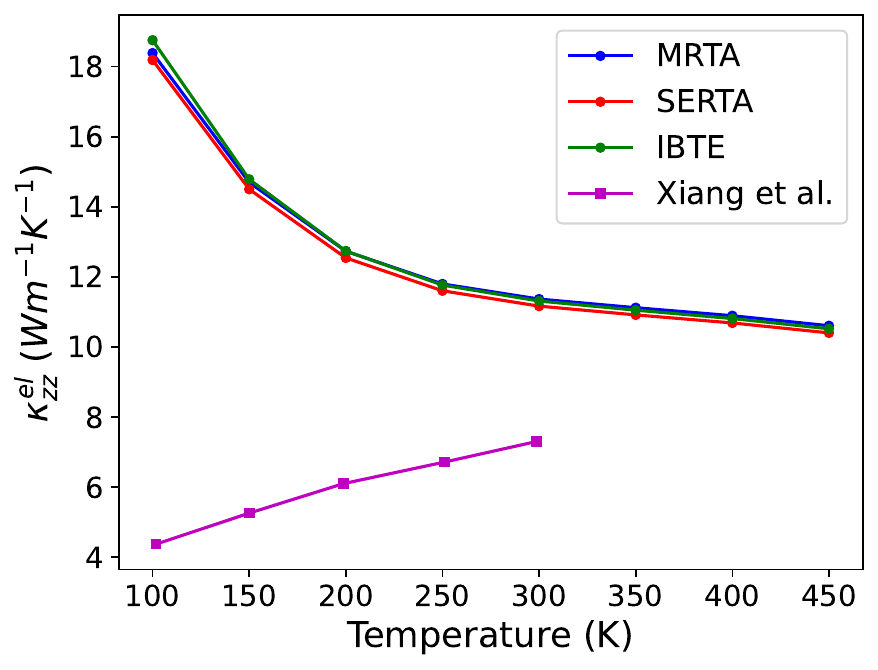}
        \caption{Temperature dependence of the xx and zz components of the electronic thermal conductivities, calculated using RTA and IBTE methods. Experimental data come from~\cite{xiang2017anisotropic}.}
        \label{fig:kappa}
\end{figure}
The results are shown in Figure~\ref{fig:kappa}. Our x values are within 30\% of the experimental results of Xiang et al.~\cite{xiang2017anisotropic}, which is significant compared to the agreement in $\sigma$. 
The comparison with experimental results, however, should be taken with a grain of salt because Ref.~\cite{xiang2017anisotropic} \emph{estimates} $\kappa^{el}$ using the Wiedemann-Franz relation, based on the measured values of $ \rho_{xx}$. 
We verify this explicitly in the following by comparing Lorenz numbers. 
The differences between the experimental and theoretical Seebeck coefficients, especially along $z$, will also carry over since $\mathbf{S}$ appears explicitly in the expression for $\boldsymbol{\kel}$ (Eq.~\ref{eq:kappa_IBTE}). 

\subsection{Peltier coefficient}

\begin{figure}[h]
     \centering
        \includegraphics[width=1\linewidth]{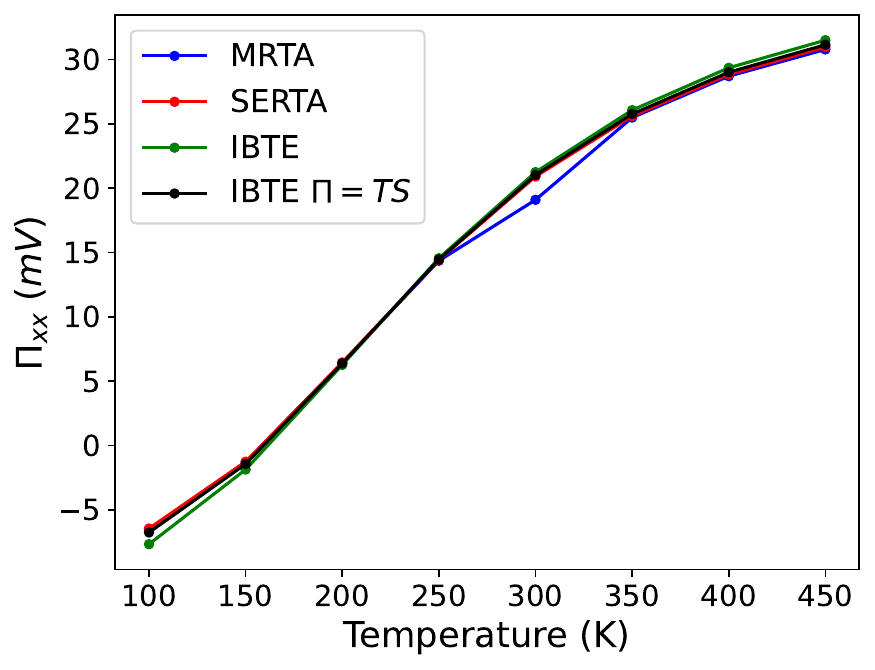}\hfill
        \includegraphics[width=1\linewidth]{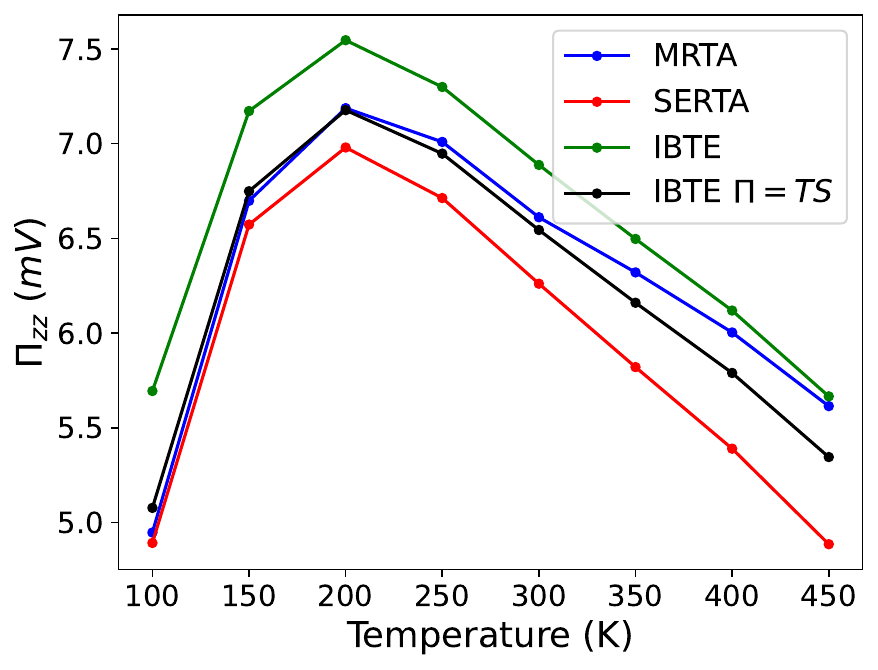}
        \caption{Temperature dependence of the xx and zz components of the Peltier coefficient, calculated using RTA and IBTE methods. We compare also the results obtained using Eq.~\ref{eq:Peltier} or using $\Pi=TS$ in IBTE. }
        \label{fig:Pi}
\end{figure}
The Peltier coefficient characterizes the amount of heat current transferred when an electric current is applied (as distinct from Joule heating). We derive its expression in the IBTE framework, imposing $\TT = 0$:
\begin{equation}
\mathbf{\Pi} = \frac{\jj^{Q}}{\jj^{e}} = \frac{\LL_{21} \cdot \EE - \LL_{22} \cdot \frac{\TT}{T}}{\LL_{11} \cdot \EE - \LL_{12} \cdot \frac{\TT}{T}} = \frac{\LL_{21}}{\LL_{11}}.
\label{eq:Peltier}
\end{equation}
Note that, in most of the thermoelectric literature, one imposes $\mathbf{\Pi} = T \mathbf{S}$ ($=  \frac{\LL_{12}}{\LL_{11}}$). This expression is valid only if time reversal symmetry is present (i.e., if there is no applied magnetic field and if the material is non-magnetic). If Eq.~\ref{eq:reci_onager} is verified (as it is the case in the RTA formalism), then the equivalence is imposed.

In Figure~\ref{fig:Pi}, we report  the calculated Peltier coefficient, and compare the IBTE values  obtained either using the full formula (Eq.~\ref{eq:Peltier}) or using $\Pi=TS$. The two considerations can differ in our case, and $\Pi_{zz}$ shows finite deviation from $TS_{zz}$. The differences could be caused by numerical effects when calculating $\LL_{12}$ and $\LL_{21}$ explicitly. 
For TaAs, we did not find any simultaneous measurements of $\mathbf{S}$ and $\mathbf{\Pi}$ in the literature, which would be a useful and stringent test of our calculations and experimental precision. 


\subsection{Lorenz number}

As anticipated above, we calculate the Lorenz number $L$ (not to be confused with the Onsager coefficients), which is the proportionality constant coming from the Wiedemann-Franz (WF) relation. $L$ reflects the ratio of the electronic thermal conductivity $\kel$ to the electrical conductivity $\sigma$. 
In a simple metal, $L$ is just proportional to the temperature $T$, with:
\begin{equation}
    L = \frac{\kel}{\sigma T}.
\end{equation}
The reference Sommerfeld value is $L_{0}= 2.445 \times 10^{-8} \ V^{2} K^{-2}$.
\begin{table*}[ht]
\centering
\begin{tabularx}{0.9\textwidth}{p{3cm}|>{\centering\arraybackslash}X>{\centering\arraybackslash}X>{\centering\arraybackslash}X>{\centering\arraybackslash}X>{\centering\arraybackslash}X>{\centering\arraybackslash}X>{\centering\arraybackslash}X>{\centering\arraybackslash}X}
$L_{xx} / L_0$ 
& 100K & 150K & 200K & 250K & 300K & 350K & 400K & 450K \\
\hline
SERTA & 1.178 & 1.575 & 1.697 & 1.603 & 1.440 & 1.288 & 1.166 & 1.063 \\
MRTA & 1.174 & 1.571 & 1.689 & 1.595 & 1.284 & 1.276 & 1.153 & 1.055 \\
IBTE & 1.174 & 1.600 & 1.718 & 1.620 & 1.448 & 1.288 & 1.157 & 1.0593 \\

\end{tabularx}

\caption{\label{tab:Lxx}Relative Lorenz number for x direction, within RTA and IBTE methods, compared to experiment~\cite{xiang2017anisotropic}.}
\end{table*}

\begin{table*}[ht]
\centering
\begin{tabularx}{0.9\textwidth}{p{3cm}|>{\centering\arraybackslash}X>{\centering\arraybackslash}X>{\centering\arraybackslash}X>{\centering\arraybackslash}X>{\centering\arraybackslash}X>{\centering\arraybackslash}X>{\centering\arraybackslash}X>{\centering\arraybackslash}X}
$L_{zz}/L_0$ 
& 100K & 150K & 200K & 250K & 300K & 350K & 400K & 450K \\
\hline
SERTA &  1.043 &   0.920 &   0.867 &   0.855 &   0.863 &   0.875 &   0.879 &   0.875 \\
MRTA &  1.039 &   0.920 &   0.867 &   0.855 &   0.838 &   0.875 &   0.879 &   0.871  \\
IBTE &  1.035 &   0.916 &   0.863 &   0.851 &   0.859 &   0.871 &   0.875 &   0.871 \\

\end{tabularx}

\caption{\label{tab:Lzz}Relative Lorenz number for z direction, within RTA and IBTE methods, compared to experiment~\cite{xiang2017anisotropic}.}
\end{table*}

In Tables~\ref{tab:Lxx} and~\ref{tab:Lzz}, we show the relative Lorenz number along both directions, and compare the different BTE results. 
As one can see, our results for $L_{xx}$ are quite far (up to 70\%) from the reference Sommerfeld value $L_{0}$, especially between 150 and 350 K, whereas the $L_{zz}$ results are closer to $L_{0}$, but lower by up to 15\%. In a topological Weyl semimetal such as TaAs, the WF ``law'' can be broken given the non-parabolic and topological electronic structure. 
We recall that Ref.~\cite{xiang2017anisotropic} presumes the validity of the WF relation approximating $L/L_0 =1$.
Part of the anisotropy they observe corresponds to the opposite trends in $L$ for the $x$ and $z$ directions.

\section{Conclusion}

We present a full first-principles study of the Weyl semimetal TaAs. Establishing its electronic structure and phonon dynamics, we evaluate the EPC and characterize the thermoelectric transport in this material by calculating the transport coefficients using two approaches for solving the BTE: the RTA and the IBTE method. We present some detail of the theoretical development, emphasizing the derivation of a double IBTE technique in electrical and thermal gradients, respectively. We show that our conductivity calculations are close to the experimental values, whereas the results for Seebeck coefficients are of the same order of magnitude. We estimate the effect of doping within the rigid band approximation, to compare with the natural doping of experimental samples. In addition, we present our calculated electronic thermal conductivities. We discuss the reciprocity of the off-diagonal Onsager coefficient in IBTE by comparing results for the Seebeck and Peltier coefficients. In our double BTE formalism $\LL_{12}$ and $\LL_{21}$ are allowed to be different, enabling calculations in systems without time-reversal symmetry.  Comparing $\Pi$ and $TS$ quantifies the numerical convergence for time-symmetric systems. Finally, we calculate the Lorenz number and show that the Wiedemann-Franz relation is not verified for Weyl semimetals such as TaAs.

\section*{acknowledgments}
The authors acknowledge important and stimulating discussions with JM Lihm about Onsager reciprocity.
We acknowledge funding by ULiege, the Fonds National de la Recherche Scientifique (FRS-FNRS Belgium) for 
the Excellence of Science (EOS) program (grant 40007563-CONNECT) and F\'ed\'eration Wallonie Bruxelles for ARC project DREAMS (G.A. 21/25-11).
MJV is partly funded by the Gravitation research program “Materials for the Quantum Age” (QuMat G.A. 024.005.006), part of the  program financed by the Dutch Ministry of Education, Culture and Science (OCW)
This work is an outcome of the Shapeable 2D magnetoelectronics by design project (SHAPEme, EOS Project No. 560400077525) that has received funding from the FWO and FRS-FNRS under the Belgian Excellence of Science (EOS) program.
Simulation time was awarded by PRACE (Optospin project id. 2020225411) on Discoverer at Sofiatech Bulgaria,
EuroHPC (Extreme grant EHPC-EXT-2023E02-050) on Marenostrum5 at BSC, Spain,
by the CECI (FRS-FNRS Belgium Grant No. 2.5020.11),
as well as the Lucia Tier-1 of the F\'ed\'eration Wallonie-Bruxelles (Walloon Region grant agreement No. 1910247).

\appendix

\section{Scattering operator}
\label{appendix:scattering_operator}

The scattering integral $\mcI_{n\kk}[f]$ includes all possible microscopic scattering events due to electron-phonon and electron-electron interaction, scattering with ionized impurities, defects, etc. 
The most commonly used form for the (one-particle) scattering integral is:
\begin{equation}
\begin{multlined}
\mcI_{n\kk}[f]  =
\sum_m \int 
\dfrac{\dd\kk'}{\Omega_{\mathrm{BZ}}} 
 f_{m\kk'}\,(1-f_\nk)\, W_{m\kk',\nk} \\
-f_\nk\,(1-f_{m\kk'})\,W_{\nk,m\kk'}
\end{multlined}
\end{equation}
where $W_{\nk,m\kk'}$ is the probability per unit time for a transition between the initial state $|\nk\rangle$
and the final state $|m\kk'\rangle$, 
and the occupation factors $f$ account for the exclusion principle. 
By definition, the Fermi-Dirac equilibrium distribution 
%
\begin{equation}
f^0(\ee) = 
\frac{1}{1+ \exp{(\frac{\ee-\mu}{KT})}}
\end{equation}
is a solution of the BTE in the absence of external fields, we thus have:
\begin{equation}
\label{eq:I=0}
\mcI_{n\kk}[f^0] = 0.
\end{equation}
%
%
%
%
%
In our case, we restrict the discussion to the EPC interaction.
This inelastic term gives the most important contribution at room temperature, and introduces a significant temperature dependence in the transport properties due to the Bose-Einstein phonon population.
The Hamiltonian
\begin{equation}
\label{eq:h_eph}
    \hat{H}_\text{e-ph} = \frac{1}{\sqrt{N_p}} \sum_{\substack{\kb,\qb \\ mn\nu}} 
                   \gkq\, \hat{c}^\dagger_{m\kb+\qb} \hat{c}_{n\kb}
                   (\hat{a}_\qnu + \hat{a}^\dagger_{-\qnu})
\end{equation}
describes the coupling to first order in the atomic displacements, with $N_p$ the number of unit cells in the Born-von K\'arm\'an supercell.
$\hat{c}^\dagger_{n\kb}$ and $\hat{c}_{n\kb}$ ($\hat{a}^\dagger_{\qnu}$ and $\hat{a}_\qnu$) are fermionic (bosonic) creation and destruction operators,
and $\gkq$ the e-ph coupling matrix elements~\cite{Giustino2017}.
The transition probability is obtained from the previous Hamiltonian using Fermi's golden rule.
The final result reads:
\begin{equation}
\begin{multlined}
\mcI_\nk^{\text{e-ph}}[f] = 
\dfrac{2 \pi}{N_\qq} \sum_{m\qq\nu} 
|\gkq|^2 \\ \times
\Bigl [
\delta(\enk - \emkq - \wqnu) 
P^- \\ + 
\delta(\enk - \emkq + \wqnu) P^+
\Bigr ]
\end{multlined}
\end{equation}
where the absorption and emission factors $P$ depend on $f$ and the 
phonon occupation $n_\qnu$ according to:
\begin{equation}
P^+  = 
-f_\nk (1 - f_\mkq) n_\qnu 
+ (1 - f_\nk) f_\mkq (n_\qnu + 1)
\end{equation}
and
\begin{equation}
P^- =  
- f_\nk (1 - f_\mkq) (n_\qnu + 1)
+ (1 - f_\nk)f_\mkq  n_\qnu.
\end{equation}
The full BTE for the steady state is a non-linear integro-differential equation. In practice, one is usually interested in the linear response of the system, so it is customary to replace the scattering integral with its linearized version, considering and using Eq.~\ref{eq:I=0}.
It is easy to show that the first order variation of the $P$ factors is given by:
\begin{equation}
\delta P^+ =
\delta f_{m\kq} (1 + n_\qnu - f^0_\nk )
-\delta f_\nk (f^0_{m\kq} + n_\qnu)
\end{equation}
\begin{equation}
\delta P^- = 
\delta f_{m\kq} (f^0_\nk + n_\qnu)
-\delta f_\nk (1 + n_\qnu - f^0_{m\kq}).
\end{equation}
The linearized e-ph scattering integral thus reads:
\begin{equation}
\begin{multlined}
 \mcJ^\eph_{n\kk}[f^0, \delta f] =
- \dfrac{\delta f_\nk}{\tau^0_\nk}
+ \dfrac{2 \pi}{N_\qq}
\sum_{m\qnu} |\gkq|^2 \\
\times \bigl [
(1 + n_\qnu - f^0_{n\kk}) \delta^+ + 
(n_\qnu + f^0_{n\kk}) \delta^- 
\bigr ]
\delta f_{m\kq} \\
= - \dfrac{\delta f_\nk}{\tau^0_\nk} + 
\mcL^\eph_{n\kk}[f^0, \delta f] 
\end{multlined}
\end{equation}
where $\tau^0_\nk$ is given by
\begin{equation}
\begin{multlined}
\dfrac{1}{\tau^0_\nk} =  
\dfrac{2\pi}{N_\qq} \sum_{m\qnu} |\gkq|^2 \\
\times \bigl [
(n_\qnu + f^0_{m\kq}) \delta^+ + 
(1 + n_\qnu - f^0_{m\kq}) \delta^- 
\bigr ] \\ 
= 2\,\Im\Sigma^{\text{e-ph}}_\nk(\ww=\ee_\nk)
\end{multlined}
\end{equation}
and is related to the imaginary part of the e-ph self-energy~\cite{Giustino2017}. 
$\tau$ represents the lifetime of a charged excitation due to e-ph scattering, and is not necessarily equal to the transport lifetime.
Note that we have introduced the following shorthand notations:
\begin{equation}
\delta^+ = \delta(\enk - \emkq + \wqnu) 
\end{equation}
\begin{equation}
\delta^- = \delta(\enk - \emkq - \wqnu).
\end{equation}
The derivation is quite lengthy and is not reported here, but more details can be found in~\cite{ponce2020first}.
There are, however, some points that are worth mentioning.
First of all, the scattering is inelastic since we are dealing with a time-dependent perturbation.
The factors $n_\qnu$ (absorption) and $1+n_\qnu$ (spontaneous plus stimulated emission) 
are typical of interaction terms as in Eq.~\ref{eq:h_eph} since the Hamiltonian can only connect two states in which the number of phonons differs by one.
Finally, we should note that $n_\qnu$ is, in 
principle, the out of equilibrium phonon distribution obtained by solving an analogous Boltzmann equation for phonons in the presence of phonon-phonon scattering (anharmonic terms) and e-ph interaction.
This means that $f$ and $n$ are the solution of two coupled Boltzmann equations~\cite{fiorentini2016}.
In many applications, however, we can, to good approximation, assume that phonons are in thermodynamic equilibrium and $n_\qnu$ is replaced by the Bose-Einstein distribution:
\begin{equation}
n^0(\ww) = 
\frac{1}{\exp{(\frac{\ww}{KT}) - 1}}.
\end{equation}
%
%


\section{Evaluation of the conductivity and Seebeck coefficient with simulated doping}
\label{section:S_doping}

In this section, we present the results gathered for the electrical conductivity and the Seebeck coefficient under simulated doping within the rigid band approximation, which consists in raising or lowering the chemical potential without modifying the previously established band structure.

\begin{figure}[h]
     \centering
        \includegraphics[width=1\linewidth]{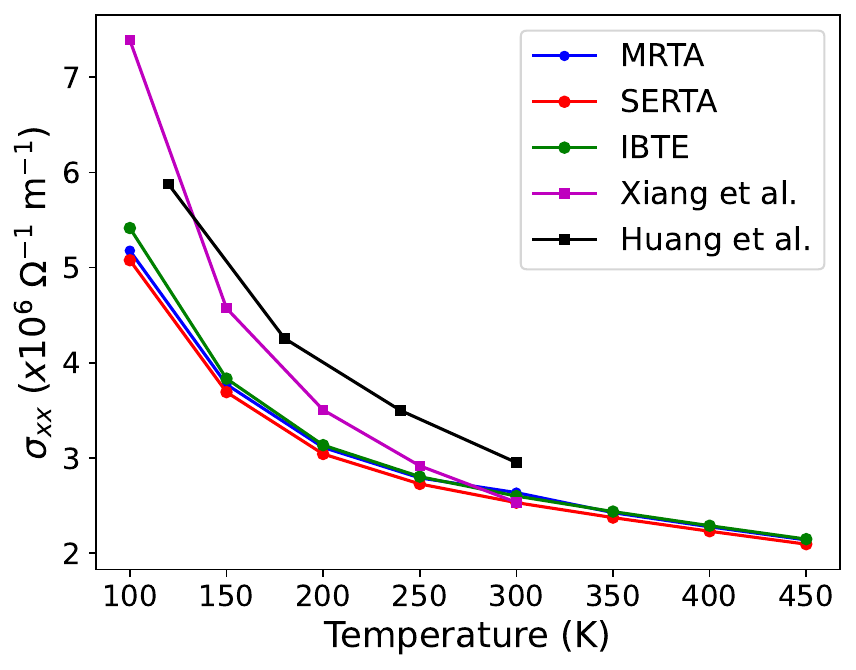}
        \includegraphics[width=1\linewidth]{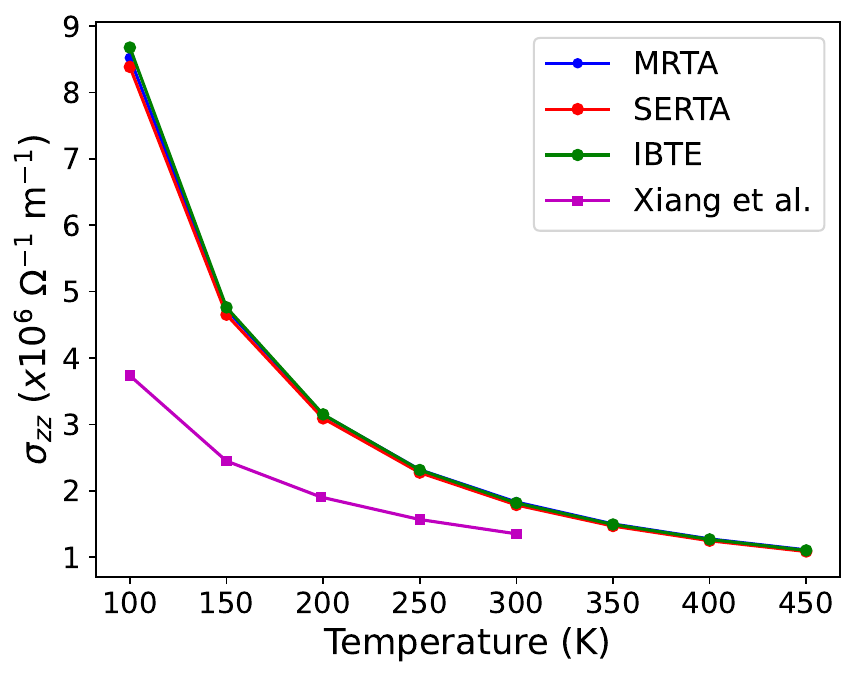}
        \caption{Temperature dependence of the xx and zz components of the electrical conductivity, calculated using RTA and IBTE methods, considering a simulated doping within the rigid band approximation of $+1 \; 10^{19}$ electronic charges per $cm^{3}$ (p-type doping). Experimental data come from~\cite{huang2015observation} and~\cite{xiang2017anisotropic}.}
        \label{fig:sigma_doped}
\end{figure}

As can be seen in Figure \ref{fig:sigma_doped}, under p-type doping of $+1 \;10^{19}$ carriers per $cm^{3}$, the electrical conductivity becomes anisotropic, but in the opposite direction to Ref.~\cite{xiang2017anisotropic}.

\begin{figure}[t]
        \vspace{1em}
        \includegraphics[width=1\linewidth]{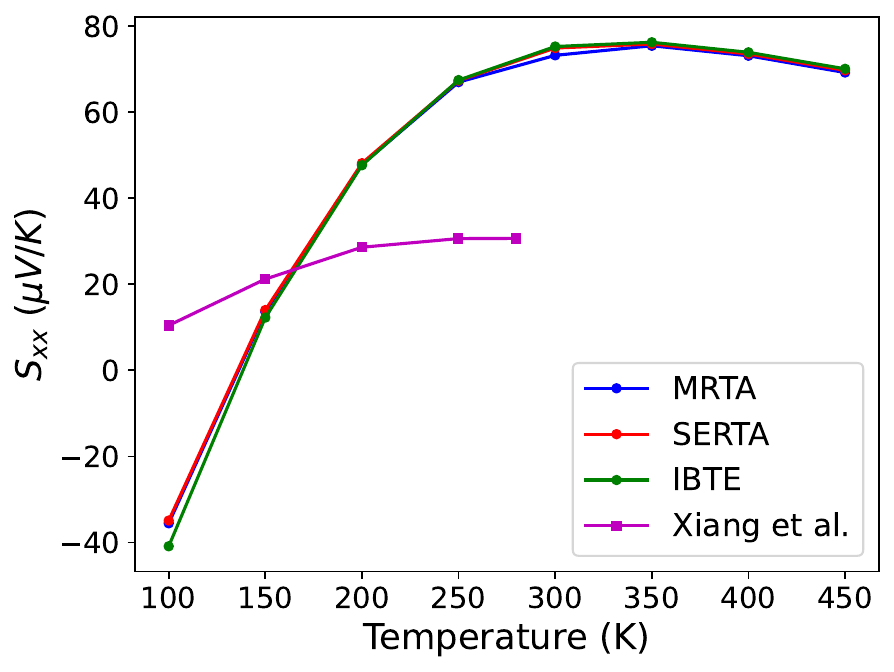}
        \caption{Temperature dependence of the x component of the Seebeck coefficient, using RTA and IBTE and considering a rigid band doping of $+1 \;10^{19}$ carriers per $cm^{3}$ (p-type). Experimental data from~\cite{xiang2017anisotropic}.}
        \label{fig:SBK_1e+19}
\end{figure}

\begin{figure}[h]
        \includegraphics[width=1\linewidth]{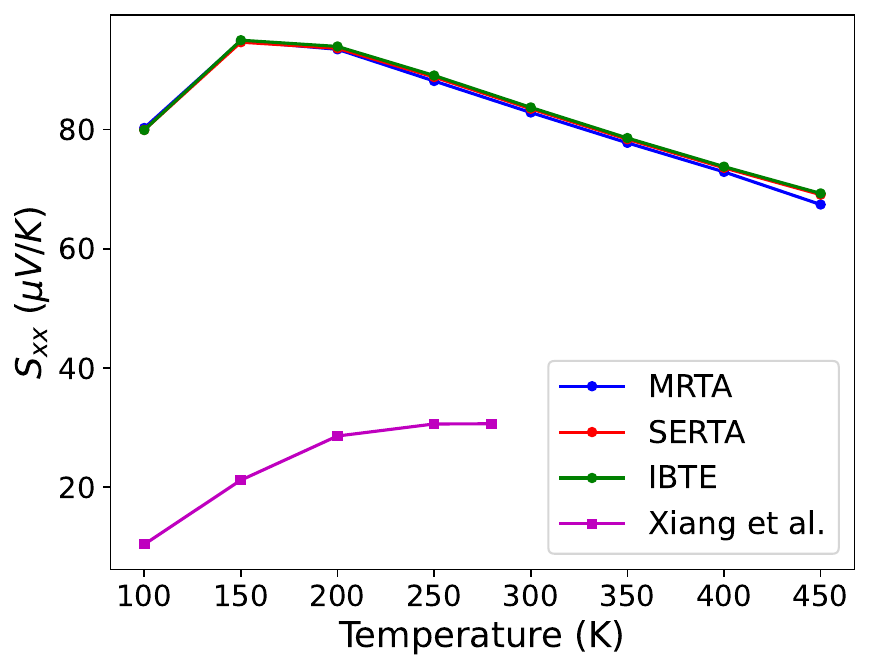}
        \caption{Temperature dependence of the x component of the Seebeck coefficient, using RTA and IBTE and considering a rigid band doping of $+5 \ 10^{19}$ carriers per $cm^{3}$ (p-type). Experimental data from~\cite{xiang2017anisotropic}.}
        \label{fig:SBK_5e+19}
\end{figure}

We also calculate the Seebeck coefficient under p-type doping. 
At $+10^{19}$ carriers per $cm^{3}$, ${S_{xx}}$ is still negative (Fig.~\ref{fig:SBK_1e+19}), but less so. While, at $+5 \ 10^{19}$ carriers per $cm^{3}$ (Fig.~\ref{fig:SBK_5e+19}), the Seebeck coefficient along the xx direction is totally positive, as also found in Ref.~\cite{peng2016high} within the cRTA.

\vspace{1em}
Rigid band doping is not sufficient to explain simultaneously the temperature variations of both $S$ and $\sigma$, suggesting there may be n and p co-doping.

\vspace{3em}

\bibliography{apssamp}

\end{document}

%% file: macros.tex
\newcommand{\kb}{\mathbf{k}}
\newcommand{\qb}{\mathbf{q}}
\newcommand{\qq}{\mathbf{q}}
\newcommand{\rb}{\mathbf{r}}
\newcommand{\Gb}{\mathbf{G}}
\newcommand{\im}{i}
\newcommand{\enk}{\varepsilon_{n\kb}}
\newcommand{\emkq}{\varepsilon_{m\kb+\qb}}
\newcommand{\ef}{\varepsilon_F}
\newcommand{\wnuq}{\omega_{\qb\nu}}
\newcommand{\qnu}{{{\qb\nu}}}
\newcommand{\wqnu}{{\omega_{\qnu}}}
\newcommand{\eqnu}{{\varepsilon_\qnu}}
\newcommand{\vnk}{\mathbf{v}_{n\kb}}
\newcommand{\vmkq}{\mathbf{v}_{m\kb+\qq}}
\newcommand{\vnka}{\mathrm{v}_{n\kb,\alpha}}
\newcommand{\vnkb}{\mathrm{v}_{n\kb,\beta}}
\newcommand{\netcdf}{\textsc{netCDF}\xspace}
\newcommand{\Lc}{\mathcal{L}}
\newcommand{\Sc}{\mathcal{S}}
\newcommand{\abinit}{\textsc{Abinit}\xspace}
\newcommand{\anaddb}{\texttt{anaddb}\xspace}
\newcommand{\qe}{\textsc{QE}\xspace}
\newcommand{\abipy}{AbiPy\xspace}
\newcommand{\epw}{\textsc{EPW}\xspace}
\newcommand{\perturbo}{\textsc{Perturbo}\xspace}
\newcommand{\boltztrap}{\textsc{Boltztrap}\xspace}
\newcommand{\yambo}{Yambo\xspace}
\newcommand{\pseudodojo}{\textsc{PseudoDojo}\xspace}
\newcommand{\gkq}{g_{mn\nu}(\kb,\qb)}
\newcommand{\gkkpL}{g_{mn\nu}^\Lc(\kb,\qb)}
\newcommand{\mrtakkp}{\alpha_{mn\kb,\qb}}
\newcommand{\abinitio}{\textit{ab initio}\xspace}
\newcommand{\grid}[1]{${#1}\times{#1}\times{#1}$}
\newcommand{\RR}{{\mathbf R}}
\newcommand{\rr}{{\mathbf r}}
\newcommand{\ie}{{\emph{i.e.}}\;}
\newcommand{\eg}{{\emph{e.g.}}\,\;}
\newcommand{\PDER}[2]{\dfrac{\partial #1}{\partial #2}}
\newcommand{\GG}{{\bf G}}
\newcommand{\GGp}{{\bf G}}
\newcommand{\qG}{{\bf q+G}}
\newcommand{\kG}{{\bf k+G}}
\newcommand{\FM}{{\text{FM}}}
\newcommand{\DW}{{\text{DW}}}
\newcommand{\KS}{{\text{KS}}}
\newcommand{\VH}{{V^{\text{H}}}}
\newcommand{\vH}{{v^{\text{H}}}}
\newcommand{\Vxc}{{V^{\text{xc}}}}
\newcommand{\vxc}{{v^{\text{xc}}}}
\newcommand{\Vscf}{{V^{\text{scf}}}}
\newcommand{\vscf}{{v^{\text{scf}}}}
\newcommand{\Vloc}{{V^{\text{loc}}}}
\newcommand{\vloc}{{v^{\text{loc}}}}
\newcommand{\Vnl}{{V^{\text{nl}}}}
\newcommand{\vnl}{{v^{\text{nl}}}}
\newcommand{\vks}{{v^\KS}}
\newcommand{\Zstar}{{\bf Z}^*}
\newcommand{\BZ}{{\text{BZ}}}
\newcommand{\IBZ}{{\text{IBZ}}}
\newcommand{\epsinf}{{\bm{\epsilon}}^\infty}
\newcommand{\zero}{\mathbf{0}}

\newcommand{\mcR}{{\mathcal{R}}}
\newcommand{\mcS}{{\mathcal{S}}}
\newcommand{\mcO}{{\mathcal{O}}}
\newcommand{\Atm}{{\bf A}}
\newcommand{\omcR}{{\hat\mcR}}
\newcommand{\omcS}{{\hat\mcS}}
\newcommand{\HH}{{\hat H}}
\newcommand{\ee}{\varepsilon}
\newcommand{\RRm}{\mcR^{-1}} 
\newcommand{\Ri}{\mcR^{-1}}
\newcommand{\Si}{\mcS^{-1}}
\newcommand{\Rit}{\mcR^{-1\tau}}
\newcommand{\mcRm}{\mcR^{-1}}
\newcommand{\btau}{{\bf \bm \tau}}
\newcommand{\bvec}{{\bf f}}
\newcommand{\rmt}{{\rr-\btau}}
\newcommand{\rmv}{{\rr-\bvec}}
\newcommand{\nk}{{n\kk}}
\newcommand{\mk}{{m\kk}}
\newcommand{\mkq}{{m\kq}}
\newcommand{\dd}{{\,\text{d}}}
\newcommand{\oRv}{{\hat\mcR_\bvec}}
\newcommand{\oSv}{{\hat\mcS_\bvec}}
\newcommand{\etal}{\textit{et al.}\xspace}
\newcommand{\dec}{{\Delta\varepsilon_c}}
\newcommand{\dev}{{\Delta\varepsilon_v}}
\newcommand{\mcB}{\mathcal{B}}
\newcommand{\mcC}{\mathcal{C}}
\newcommand{\mcE}{\mathcal{E}}
\newcommand{\mcF}{\mathcal{F}}
\newcommand{\mcG}{\mathcal{G}}
\newcommand{\mcV}{\mathcal{V}}
\newcommand{\mcI}{\mathcal{I}}
\newcommand{\mcJ}{\mathcal{J}}
\newcommand{\mcL}{\mathcal{L}}
\newcommand{\wbz}{\Omega_\BZ}
\newcommand{\dkbz}{{\dfrac{\dd\kk}{\wbz}}}

\newcommand{\kk}{\mathbf{k}}
\newcommand{\kq}{\mathbf{k + q}}
\newcommand{\jj}{\mathbf{j}}
\newcommand{\vv}{\mathbf{v}}
\newcommand{\EE}{\mathbf{E}}
\newcommand{\bmcE}{{\mathbf{\mcE}}}
\newcommand{\bmcF}{{\mathbf{\mcF}}}
\newcommand{\FF}{\mathbf{F}}
\newcommand{\BB}{\mathbf{B}}
\newcommand{\TT}{\boldsymbol{\nabla}T}
\newcommand{\ww}{\omega}
\newcommand{\ab}{{\alpha\beta}}
\newcommand{\bme}{{\bm{\epsilon}}}
\newcommand{\dfodenk}{{\PDER{f^0}{\ee_\nk}}}
\newcommand{\dkopi}{{\dfrac{\dd\kk}{\;(2\pi)^3}}}
\newcommand{\eph}{{\text{e-ph}}}
\newcommand{\Ieph}{\mcI^\eph_{n\kk}[f^0, \delta f]}
\newcommand{\Jeph}{\mcJ^\eph_{n\kk}[f^0, \delta f]}
\newcommand{\Leph}{\mcL^\eph_{n\kk}[f^0, \delta f]}
\newcommand{\Lnk}{\mcL_{n\kk}[f^0, \delta f]}
\newcommand{\LL}{\mathbf{L}}
\newcommand{\kel}{\mathbf{\kappa}^{el}}

%% file: main.bbl
\begin{thebibliography}{32}%
\makeatletter
\providecommand \@ifxundefined [1]{%
 \@ifx{#1\undefined}
}%
\providecommand \@ifnum [1]{%
 \ifnum #1\expandafter \@firstoftwo
 \else \expandafter \@secondoftwo
 \fi
}%
\providecommand \@ifx [1]{%
 \ifx #1\expandafter \@firstoftwo
 \else \expandafter \@secondoftwo
 \fi
}%
\providecommand \natexlab [1]{#1}%
\providecommand \enquote  [1]{``#1''}%
\providecommand \bibnamefont  [1]{#1}%
\providecommand \bibfnamefont [1]{#1}%
\providecommand \citenamefont [1]{#1}%
\providecommand \href@noop [0]{\@secondoftwo}%
\providecommand \href [0]{\begingroup \@sanitize@url \@href}%
\providecommand \@href[1]{\@@startlink{#1}\@@href}%
\providecommand \@@href[1]{\endgroup#1\@@endlink}%
\providecommand \@sanitize@url [0]{\catcode `\\12\catcode `\$12\catcode
  `\&12\catcode `\#12\catcode `\^12\catcode `\_12\catcode `\%12\relax}%
\providecommand \@@startlink[1]{}%
\providecommand \@@endlink[0]{}%
\providecommand \url  [0]{\begingroup\@sanitize@url \@url }%
\providecommand \@url [1]{\endgroup\@href {#1}{\urlprefix }}%
\providecommand \urlprefix  [0]{URL }%
\providecommand \Eprint [0]{\href }%
\providecommand \doibase [0]{https://doi.org/}%
\providecommand \selectlanguage [0]{\@gobble}%
\providecommand \bibinfo  [0]{\@secondoftwo}%
\providecommand \bibfield  [0]{\@secondoftwo}%
\providecommand \translation [1]{[#1]}%
\providecommand \BibitemOpen [0]{}%
\providecommand \bibitemStop [0]{}%
\providecommand \bibitemNoStop [0]{.\EOS\space}%
\providecommand \EOS [0]{\spacefactor3000\relax}%
\providecommand \BibitemShut  [1]{\csname bibitem#1\endcsname}%
\let\auto@bib@innerbib\@empty
\bibitem [{\citenamefont {Huang}\ \emph {et~al.}(2015)\citenamefont {Huang},
  \citenamefont {Zhao}, \citenamefont {Long}, \citenamefont {Wang},
  \citenamefont {Chen}, \citenamefont {Yang}, \citenamefont {Liang},
  \citenamefont {Xue}, \citenamefont {Weng}, \citenamefont {Fang} \emph
  {et~al.}}]{huang2015observation}%
  \BibitemOpen
  \bibfield  {author} {\bibinfo {author} {\bibfnamefont {X.}~\bibnamefont
  {Huang}}, \bibinfo {author} {\bibfnamefont {L.}~\bibnamefont {Zhao}},
  \bibinfo {author} {\bibfnamefont {Y.}~\bibnamefont {Long}}, \bibinfo {author}
  {\bibfnamefont {P.}~\bibnamefont {Wang}}, \bibinfo {author} {\bibfnamefont
  {D.}~\bibnamefont {Chen}}, \bibinfo {author} {\bibfnamefont {Z.}~\bibnamefont
  {Yang}}, \bibinfo {author} {\bibfnamefont {H.}~\bibnamefont {Liang}},
  \bibinfo {author} {\bibfnamefont {M.}~\bibnamefont {Xue}}, \bibinfo {author}
  {\bibfnamefont {H.}~\bibnamefont {Weng}}, \bibinfo {author} {\bibfnamefont
  {Z.}~\bibnamefont {Fang}}, \emph {et~al.},\ }\bibfield  {title} {\bibinfo
  {title} {Observation of the chiral-anomaly-induced negative magnetoresistance
  in 3d weyl semimetal taas},\ }\href@noop {} {\bibfield  {journal} {\bibinfo
  {journal} {Physical Review X}\ }\textbf {\bibinfo {volume} {5}},\ \bibinfo
  {pages} {031023} (\bibinfo {year} {2015})}\BibitemShut {NoStop}%
\bibitem [{\citenamefont {Lv}\ \emph {et~al.}(2015)\citenamefont {Lv},
  \citenamefont {Weng}, \citenamefont {Fu}, \citenamefont {Wang}, \citenamefont
  {Miao}, \citenamefont {Ma}, \citenamefont {Richard}, \citenamefont {Huang},
  \citenamefont {Zhao}, \citenamefont {Chen} \emph
  {et~al.}}]{lv2015experimental}%
  \BibitemOpen
  \bibfield  {author} {\bibinfo {author} {\bibfnamefont {B.}~\bibnamefont
  {Lv}}, \bibinfo {author} {\bibfnamefont {H.}~\bibnamefont {Weng}}, \bibinfo
  {author} {\bibfnamefont {B.}~\bibnamefont {Fu}}, \bibinfo {author}
  {\bibfnamefont {X.~P.}\ \bibnamefont {Wang}}, \bibinfo {author}
  {\bibfnamefont {H.}~\bibnamefont {Miao}}, \bibinfo {author} {\bibfnamefont
  {J.}~\bibnamefont {Ma}}, \bibinfo {author} {\bibfnamefont {P.}~\bibnamefont
  {Richard}}, \bibinfo {author} {\bibfnamefont {X.}~\bibnamefont {Huang}},
  \bibinfo {author} {\bibfnamefont {L.}~\bibnamefont {Zhao}}, \bibinfo {author}
  {\bibfnamefont {G.}~\bibnamefont {Chen}}, \emph {et~al.},\ }\bibfield
  {title} {\bibinfo {title} {Experimental discovery of weyl semimetal taas},\
  }\href@noop {} {\bibfield  {journal} {\bibinfo  {journal} {Physical Review
  X}\ }\textbf {\bibinfo {volume} {5}},\ \bibinfo {pages} {031013} (\bibinfo
  {year} {2015})}\BibitemShut {NoStop}%
\bibitem [{\citenamefont {Xiang}\ \emph {et~al.}(2017)\citenamefont {Xiang},
  \citenamefont {Hu}, \citenamefont {Lv}, \citenamefont {Zhang}, \citenamefont
  {Zhao}, \citenamefont {Chen}, \citenamefont {Li}, \citenamefont {Chen},\ and\
  \citenamefont {Sun}}]{xiang2017anisotropic}%
  \BibitemOpen
  \bibfield  {author} {\bibinfo {author} {\bibfnamefont {J.}~\bibnamefont
  {Xiang}}, \bibinfo {author} {\bibfnamefont {S.}~\bibnamefont {Hu}}, \bibinfo
  {author} {\bibfnamefont {M.}~\bibnamefont {Lv}}, \bibinfo {author}
  {\bibfnamefont {J.}~\bibnamefont {Zhang}}, \bibinfo {author} {\bibfnamefont
  {H.}~\bibnamefont {Zhao}}, \bibinfo {author} {\bibfnamefont {G.}~\bibnamefont
  {Chen}}, \bibinfo {author} {\bibfnamefont {W.}~\bibnamefont {Li}}, \bibinfo
  {author} {\bibfnamefont {Z.}~\bibnamefont {Chen}},\ and\ \bibinfo {author}
  {\bibfnamefont {P.}~\bibnamefont {Sun}},\ }\bibfield  {title} {\bibinfo
  {title} {Anisotropic thermal and electrical transport of weyl semimetal
  taas},\ }\href@noop {} {\bibfield  {journal} {\bibinfo  {journal} {Journal of
  Physics: Condensed Matter}\ }\textbf {\bibinfo {volume} {29}},\ \bibinfo
  {pages} {485501} (\bibinfo {year} {2017})}\BibitemShut {NoStop}%
\bibitem [{\citenamefont {Xu}\ \emph {et~al.}(2021)\citenamefont {Xu},
  \citenamefont {Liu}, \citenamefont {Seyfarth}, \citenamefont {Pourret},
  \citenamefont {Ma}, \citenamefont {Zhou}, \citenamefont {Wang}, \citenamefont
  {Qu},\ and\ \citenamefont {Jia}}]{xu2021thermoelectric}%
  \BibitemOpen
  \bibfield  {author} {\bibinfo {author} {\bibfnamefont {X.}~\bibnamefont
  {Xu}}, \bibinfo {author} {\bibfnamefont {Y.}~\bibnamefont {Liu}}, \bibinfo
  {author} {\bibfnamefont {G.}~\bibnamefont {Seyfarth}}, \bibinfo {author}
  {\bibfnamefont {A.}~\bibnamefont {Pourret}}, \bibinfo {author} {\bibfnamefont
  {W.}~\bibnamefont {Ma}}, \bibinfo {author} {\bibfnamefont {H.}~\bibnamefont
  {Zhou}}, \bibinfo {author} {\bibfnamefont {G.}~\bibnamefont {Wang}}, \bibinfo
  {author} {\bibfnamefont {Z.}~\bibnamefont {Qu}},\ and\ \bibinfo {author}
  {\bibfnamefont {S.}~\bibnamefont {Jia}},\ }\bibfield  {title} {\bibinfo
  {title} {Thermoelectric transport and phonon drag in weyl semimetal
  monochalcogenides},\ }\href@noop {} {\bibfield  {journal} {\bibinfo
  {journal} {Physical Review B}\ }\textbf {\bibinfo {volume} {104}},\ \bibinfo
  {pages} {115164} (\bibinfo {year} {2021})}\BibitemShut {NoStop}%
\bibitem [{\citenamefont {Garcia}\ \emph {et~al.}(2020)\citenamefont {Garcia},
  \citenamefont {Coulter},\ and\ \citenamefont
  {Narang}}]{garcia2020optoelectronic}%
  \BibitemOpen
  \bibfield  {author} {\bibinfo {author} {\bibfnamefont {C.~A.}\ \bibnamefont
  {Garcia}}, \bibinfo {author} {\bibfnamefont {J.}~\bibnamefont {Coulter}},\
  and\ \bibinfo {author} {\bibfnamefont {P.}~\bibnamefont {Narang}},\
  }\bibfield  {title} {\bibinfo {title} {Optoelectronic response of the type-i
  weyl semimetals taas and nbas from first principles},\ }\href@noop {}
  {\bibfield  {journal} {\bibinfo  {journal} {Physical Review Research}\
  }\textbf {\bibinfo {volume} {2}},\ \bibinfo {pages} {013073} (\bibinfo {year}
  {2020})}\BibitemShut {NoStop}%
\bibitem [{\citenamefont {Coulter}\ \emph {et~al.}(2019)\citenamefont
  {Coulter}, \citenamefont {Osterhoudt}, \citenamefont {Garcia}, \citenamefont
  {Wang}, \citenamefont {Plisson}, \citenamefont {Shen}, \citenamefont {Ni},
  \citenamefont {Burch},\ and\ \citenamefont {Narang}}]{coulter2019uncovering}%
  \BibitemOpen
  \bibfield  {author} {\bibinfo {author} {\bibfnamefont {J.}~\bibnamefont
  {Coulter}}, \bibinfo {author} {\bibfnamefont {G.~B.}\ \bibnamefont
  {Osterhoudt}}, \bibinfo {author} {\bibfnamefont {C.~A.}\ \bibnamefont
  {Garcia}}, \bibinfo {author} {\bibfnamefont {Y.}~\bibnamefont {Wang}},
  \bibinfo {author} {\bibfnamefont {V.~M.}\ \bibnamefont {Plisson}}, \bibinfo
  {author} {\bibfnamefont {B.}~\bibnamefont {Shen}}, \bibinfo {author}
  {\bibfnamefont {N.}~\bibnamefont {Ni}}, \bibinfo {author} {\bibfnamefont
  {K.~S.}\ \bibnamefont {Burch}},\ and\ \bibinfo {author} {\bibfnamefont
  {P.}~\bibnamefont {Narang}},\ }\bibfield  {title} {\bibinfo {title}
  {Uncovering electron-phonon scattering and phonon dynamics in type-i weyl
  semimetals},\ }\href@noop {} {\bibfield  {journal} {\bibinfo  {journal}
  {Physical Review B}\ }\textbf {\bibinfo {volume} {100}},\ \bibinfo {pages}
  {220301} (\bibinfo {year} {2019})}\BibitemShut {NoStop}%
\bibitem [{\citenamefont {Peng}\ \emph {et~al.}(2016)\citenamefont {Peng},
  \citenamefont {Zhang}, \citenamefont {Shao}, \citenamefont {Lu},
  \citenamefont {Zhang},\ and\ \citenamefont {Zhu}}]{peng2016high}%
  \BibitemOpen
  \bibfield  {author} {\bibinfo {author} {\bibfnamefont {B.}~\bibnamefont
  {Peng}}, \bibinfo {author} {\bibfnamefont {H.}~\bibnamefont {Zhang}},
  \bibinfo {author} {\bibfnamefont {H.}~\bibnamefont {Shao}}, \bibinfo {author}
  {\bibfnamefont {H.}~\bibnamefont {Lu}}, \bibinfo {author} {\bibfnamefont
  {D.~W.}\ \bibnamefont {Zhang}},\ and\ \bibinfo {author} {\bibfnamefont
  {H.}~\bibnamefont {Zhu}},\ }\bibfield  {title} {\bibinfo {title} {High
  thermoelectric performance of weyl semimetal taas},\ }\href@noop {}
  {\bibfield  {journal} {\bibinfo  {journal} {Nano Energy}\ }\textbf {\bibinfo
  {volume} {30}},\ \bibinfo {pages} {225} (\bibinfo {year} {2016})}\BibitemShut
  {NoStop}%
\bibitem [{\citenamefont {Xu}\ \emph {et~al.}(2020)\citenamefont {Xu},
  \citenamefont {Di~Gennaro},\ and\ \citenamefont
  {Verstraete}}]{xu2020thermoelectric}%
  \BibitemOpen
  \bibfield  {author} {\bibinfo {author} {\bibfnamefont {B.}~\bibnamefont
  {Xu}}, \bibinfo {author} {\bibfnamefont {M.}~\bibnamefont {Di~Gennaro}},\
  and\ \bibinfo {author} {\bibfnamefont {M.~J.}\ \bibnamefont {Verstraete}},\
  }\bibfield  {title} {\bibinfo {title} {Thermoelectric properties of elemental
  metals from first-principles electron-phonon coupling},\ }\href@noop {}
  {\bibfield  {journal} {\bibinfo  {journal} {Physical Review B}\ }\textbf
  {\bibinfo {volume} {102}},\ \bibinfo {pages} {155128} (\bibinfo {year}
  {2020})}\BibitemShut {NoStop}%
\bibitem [{\citenamefont {Gonze}\ \emph {et~al.}(2020)\citenamefont {Gonze},
  \citenamefont {Amadon}, \citenamefont {Antonius}, \citenamefont {Arnardi},
  \citenamefont {Baguet}, \citenamefont {Beuken}, \citenamefont {Bieder},
  \citenamefont {Bottin}, \citenamefont {Bouchet}, \citenamefont {Bousquet},
  \citenamefont {Brouwer}, \citenamefont {Bruneval}, \citenamefont {Brunin},
  \citenamefont {Cavignac}, \citenamefont {Charraud}, \citenamefont {Chen},
  \citenamefont {Côté}, \citenamefont {Cottenier}, \citenamefont {Denier},
  \citenamefont {Geneste}, \citenamefont {Ghosez}, \citenamefont {Giantomassi},
  \citenamefont {Gillet}, \citenamefont {Gingras}, \citenamefont {Hamann},
  \citenamefont {Hautier}, \citenamefont {He}, \citenamefont {Helbig},
  \citenamefont {Holzwarth}, \citenamefont {Jia}, \citenamefont {Jollet},
  \citenamefont {Lafargue-Dit-Hauret}, \citenamefont {Lejaeghere},
  \citenamefont {Marques}, \citenamefont {Martin}, \citenamefont {Martins},
  \citenamefont {Miranda}, \citenamefont {Naccarato}, \citenamefont {Persson},
  \citenamefont {Petretto}, \citenamefont {Planes}, \citenamefont {Pouillon},
  \citenamefont {Prokhorenko}, \citenamefont {Ricci}, \citenamefont
  {Rignanese}, \citenamefont {Romero}, \citenamefont {Schmitt}, \citenamefont
  {Torrent}, \citenamefont {van Setten}, \citenamefont {Troeye}, \citenamefont
  {Verstraete}, \citenamefont {Zérah},\ and\ \citenamefont
  {Zwanziger}}]{Gonze2020}%
  \BibitemOpen
  \bibfield  {author} {\bibinfo {author} {\bibfnamefont {X.}~\bibnamefont
  {Gonze}}, \bibinfo {author} {\bibfnamefont {B.}~\bibnamefont {Amadon}},
  \bibinfo {author} {\bibfnamefont {G.}~\bibnamefont {Antonius}}, \bibinfo
  {author} {\bibfnamefont {F.}~\bibnamefont {Arnardi}}, \bibinfo {author}
  {\bibfnamefont {L.}~\bibnamefont {Baguet}}, \bibinfo {author} {\bibfnamefont
  {J.-M.}\ \bibnamefont {Beuken}}, \bibinfo {author} {\bibfnamefont
  {J.}~\bibnamefont {Bieder}}, \bibinfo {author} {\bibfnamefont
  {F.}~\bibnamefont {Bottin}}, \bibinfo {author} {\bibfnamefont
  {J.}~\bibnamefont {Bouchet}}, \bibinfo {author} {\bibfnamefont
  {E.}~\bibnamefont {Bousquet}}, \bibinfo {author} {\bibfnamefont
  {N.}~\bibnamefont {Brouwer}}, \bibinfo {author} {\bibfnamefont
  {F.}~\bibnamefont {Bruneval}}, \bibinfo {author} {\bibfnamefont
  {G.}~\bibnamefont {Brunin}}, \bibinfo {author} {\bibfnamefont
  {T.}~\bibnamefont {Cavignac}}, \bibinfo {author} {\bibfnamefont {J.-B.}\
  \bibnamefont {Charraud}}, \bibinfo {author} {\bibfnamefont {W.}~\bibnamefont
  {Chen}}, \bibinfo {author} {\bibfnamefont {M.}~\bibnamefont {Côté}},
  \bibinfo {author} {\bibfnamefont {S.}~\bibnamefont {Cottenier}}, \bibinfo
  {author} {\bibfnamefont {J.}~\bibnamefont {Denier}}, \bibinfo {author}
  {\bibfnamefont {G.}~\bibnamefont {Geneste}}, \bibinfo {author} {\bibfnamefont
  {P.}~\bibnamefont {Ghosez}}, \bibinfo {author} {\bibfnamefont
  {M.}~\bibnamefont {Giantomassi}}, \bibinfo {author} {\bibfnamefont
  {Y.}~\bibnamefont {Gillet}}, \bibinfo {author} {\bibfnamefont
  {O.}~\bibnamefont {Gingras}}, \bibinfo {author} {\bibfnamefont {D.~R.}\
  \bibnamefont {Hamann}}, \bibinfo {author} {\bibfnamefont {G.}~\bibnamefont
  {Hautier}}, \bibinfo {author} {\bibfnamefont {X.}~\bibnamefont {He}},
  \bibinfo {author} {\bibfnamefont {N.}~\bibnamefont {Helbig}}, \bibinfo
  {author} {\bibfnamefont {N.}~\bibnamefont {Holzwarth}}, \bibinfo {author}
  {\bibfnamefont {Y.}~\bibnamefont {Jia}}, \bibinfo {author} {\bibfnamefont
  {F.}~\bibnamefont {Jollet}}, \bibinfo {author} {\bibfnamefont
  {W.}~\bibnamefont {Lafargue-Dit-Hauret}}, \bibinfo {author} {\bibfnamefont
  {K.}~\bibnamefont {Lejaeghere}}, \bibinfo {author} {\bibfnamefont {M.~A.~L.}\
  \bibnamefont {Marques}}, \bibinfo {author} {\bibfnamefont {A.}~\bibnamefont
  {Martin}}, \bibinfo {author} {\bibfnamefont {C.}~\bibnamefont {Martins}},
  \bibinfo {author} {\bibfnamefont {H.~P.~C.}\ \bibnamefont {Miranda}},
  \bibinfo {author} {\bibfnamefont {F.}~\bibnamefont {Naccarato}}, \bibinfo
  {author} {\bibfnamefont {K.}~\bibnamefont {Persson}}, \bibinfo {author}
  {\bibfnamefont {G.}~\bibnamefont {Petretto}}, \bibinfo {author}
  {\bibfnamefont {V.}~\bibnamefont {Planes}}, \bibinfo {author} {\bibfnamefont
  {Y.}~\bibnamefont {Pouillon}}, \bibinfo {author} {\bibfnamefont
  {S.}~\bibnamefont {Prokhorenko}}, \bibinfo {author} {\bibfnamefont
  {F.}~\bibnamefont {Ricci}}, \bibinfo {author} {\bibfnamefont {G.-M.}\
  \bibnamefont {Rignanese}}, \bibinfo {author} {\bibfnamefont {A.~H.}\
  \bibnamefont {Romero}}, \bibinfo {author} {\bibfnamefont {M.~M.}\
  \bibnamefont {Schmitt}}, \bibinfo {author} {\bibfnamefont {M.}~\bibnamefont
  {Torrent}}, \bibinfo {author} {\bibfnamefont {M.~J.}\ \bibnamefont {van
  Setten}}, \bibinfo {author} {\bibfnamefont {B.~V.}\ \bibnamefont {Troeye}},
  \bibinfo {author} {\bibfnamefont {M.~J.}\ \bibnamefont {Verstraete}},
  \bibinfo {author} {\bibfnamefont {G.}~\bibnamefont {Zérah}},\ and\ \bibinfo
  {author} {\bibfnamefont {J.~W.}\ \bibnamefont {Zwanziger}},\ }\bibfield
  {title} {\bibinfo {title} {The abinit project: Impact, environment and recent
  developments},\ }\href {https://doi.org/10.1016/j.cpc.2019.107042} {\bibfield
   {journal} {\bibinfo  {journal} {Comput. Phys. Commun.}\ }\textbf {\bibinfo
  {volume} {248}},\ \bibinfo {pages} {107042} (\bibinfo {year}
  {2020})}\BibitemShut {NoStop}%
\bibitem [{\citenamefont {Hohenberg}\ and\ \citenamefont
  {Kohn}(1964)}]{Hohenberg1964}%
  \BibitemOpen
  \bibfield  {author} {\bibinfo {author} {\bibfnamefont {P.}~\bibnamefont
  {Hohenberg}}\ and\ \bibinfo {author} {\bibfnamefont {W.}~\bibnamefont
  {Kohn}},\ }\bibfield  {title} {\bibinfo {title} {Inhomogeneous electron
  gas},\ }\href {https://doi.org/10.1103/physrev.136.b864} {\bibfield
  {journal} {\bibinfo  {journal} {Phys. Rev.}\ }\textbf {\bibinfo {volume}
  {136}},\ \bibinfo {pages} {B864} (\bibinfo {year} {1964})}\BibitemShut
  {NoStop}%
\bibitem [{\citenamefont {Kohn}\ and\ \citenamefont {Sham}(1965)}]{Kohn1965}%
  \BibitemOpen
  \bibfield  {author} {\bibinfo {author} {\bibfnamefont {W.}~\bibnamefont
  {Kohn}}\ and\ \bibinfo {author} {\bibfnamefont {L.~J.}\ \bibnamefont
  {Sham}},\ }\bibfield  {title} {\bibinfo {title} {Self-consistent equations
  including exchange and correlation effects},\ }\href
  {https://doi.org/10.1103/physrev.140.a1133} {\bibfield  {journal} {\bibinfo
  {journal} {Phys. Rev.}\ }\textbf {\bibinfo {volume} {140}},\ \bibinfo {pages}
  {A1133} (\bibinfo {year} {1965})}\BibitemShut {NoStop}%
\bibitem [{\citenamefont {Gonze}(1997)}]{Gonze1997}%
  \BibitemOpen
  \bibfield  {author} {\bibinfo {author} {\bibfnamefont {X.}~\bibnamefont
  {Gonze}},\ }\bibfield  {title} {\bibinfo {title} {First-principles responses
  of solids to atomic displacements and homogeneous electric fields:
  {Implementation} of a conjugate-gradient algorithm},\ }\href
  {https://doi.org/10.1103/physrevb.55.10337} {\bibfield  {journal} {\bibinfo
  {journal} {Phys. Rev. B}\ }\textbf {\bibinfo {volume} {55}},\ \bibinfo
  {pages} {10337} (\bibinfo {year} {1997})}\BibitemShut {NoStop}%
\bibitem [{\citenamefont {Baroni}\ \emph {et~al.}(2001)\citenamefont {Baroni},
  \citenamefont {de~Gironcoli}, \citenamefont {Dal~Corso},\ and\ \citenamefont
  {Giannozzi}}]{Baroni2001}%
  \BibitemOpen
  \bibfield  {author} {\bibinfo {author} {\bibfnamefont {S.}~\bibnamefont
  {Baroni}}, \bibinfo {author} {\bibfnamefont {S.}~\bibnamefont
  {de~Gironcoli}}, \bibinfo {author} {\bibfnamefont {A.}~\bibnamefont
  {Dal~Corso}},\ and\ \bibinfo {author} {\bibfnamefont {P.}~\bibnamefont
  {Giannozzi}},\ }\bibfield  {title} {\bibinfo {title} {Phonons and related
  crystal properties from density-functional perturbation theory},\ }\href
  {https://doi.org/10.1103/revmodphys.73.515} {\bibfield  {journal} {\bibinfo
  {journal} {Rev. Mod. Phys.}\ }\textbf {\bibinfo {volume} {73}},\ \bibinfo
  {pages} {515} (\bibinfo {year} {2001})}\BibitemShut {NoStop}%
\bibitem [{\citenamefont {Perdew}\ \emph {et~al.}(1996)\citenamefont {Perdew},
  \citenamefont {Burke},\ and\ \citenamefont {Ernzerhof}}]{PBE1}%
  \BibitemOpen
  \bibfield  {author} {\bibinfo {author} {\bibfnamefont {J.~P.}\ \bibnamefont
  {Perdew}}, \bibinfo {author} {\bibfnamefont {K.}~\bibnamefont {Burke}},\ and\
  \bibinfo {author} {\bibfnamefont {M.}~\bibnamefont {Ernzerhof}},\ }\bibfield
  {title} {\bibinfo {title} {Generalized gradient approximation made simple},\
  }\href {https://doi.org/10.1103/PhysRevLett.77.3865} {\bibfield  {journal}
  {\bibinfo  {journal} {Phys. Rev. Lett.}\ }\textbf {\bibinfo {volume} {77}},\
  \bibinfo {pages} {3865} (\bibinfo {year} {1996})}\BibitemShut {NoStop}%
\bibitem [{\citenamefont {Hamann}(2013)}]{Hamann2013}%
  \BibitemOpen
  \bibfield  {author} {\bibinfo {author} {\bibfnamefont {D.~R.}\ \bibnamefont
  {Hamann}},\ }\bibfield  {title} {\bibinfo {title} {Optimized norm-conserving
  vanderbilt pseudopotentials},\ }\href
  {https://doi.org/10.1103/physrevb.88.085117} {\bibfield  {journal} {\bibinfo
  {journal} {Phys. Rev. B}\ }\textbf {\bibinfo {volume} {88}},\ \bibinfo
  {pages} {085117} (\bibinfo {year} {2013})}\BibitemShut {NoStop}%
\bibitem [{\citenamefont {{van Setten}}\ \emph {et~al.}(2018)\citenamefont
  {{van Setten}}, \citenamefont {Giantomassi}, \citenamefont {Bousquet},
  \citenamefont {Verstraete}, \citenamefont {Hamann}, \citenamefont {Gonze},\
  and\ \citenamefont {Rignanese}}]{VANSETTEN201839}%
  \BibitemOpen
  \bibfield  {author} {\bibinfo {author} {\bibfnamefont {M.}~\bibnamefont {{van
  Setten}}}, \bibinfo {author} {\bibfnamefont {M.}~\bibnamefont {Giantomassi}},
  \bibinfo {author} {\bibfnamefont {E.}~\bibnamefont {Bousquet}}, \bibinfo
  {author} {\bibfnamefont {M.}~\bibnamefont {Verstraete}}, \bibinfo {author}
  {\bibfnamefont {D.}~\bibnamefont {Hamann}}, \bibinfo {author} {\bibfnamefont
  {X.}~\bibnamefont {Gonze}},\ and\ \bibinfo {author} {\bibfnamefont {G.-M.}\
  \bibnamefont {Rignanese}},\ }\bibfield  {title} {\bibinfo {title} {The
  pseudodojo: Training and grading a 85 element optimized norm-conserving
  pseudopotential table},\ }\href
  {https://doi.org/https://doi.org/10.1016/j.cpc.2018.01.012} {\bibfield
  {journal} {\bibinfo  {journal} {Computer Physics Communications}\ }\textbf
  {\bibinfo {volume} {226}},\ \bibinfo {pages} {39} (\bibinfo {year}
  {2018})}\BibitemShut {NoStop}%
\bibitem [{\citenamefont {Chang}\ \emph {et~al.}(2016)\citenamefont {Chang},
  \citenamefont {Liu}, \citenamefont {Rao}, \citenamefont {Wang}, \citenamefont
  {Sun},\ and\ \citenamefont {Jia}}]{chang2016phonon}%
  \BibitemOpen
  \bibfield  {author} {\bibinfo {author} {\bibfnamefont {D.}~\bibnamefont
  {Chang}}, \bibinfo {author} {\bibfnamefont {Y.}~\bibnamefont {Liu}}, \bibinfo
  {author} {\bibfnamefont {F.}~\bibnamefont {Rao}}, \bibinfo {author}
  {\bibfnamefont {F.}~\bibnamefont {Wang}}, \bibinfo {author} {\bibfnamefont
  {Q.}~\bibnamefont {Sun}},\ and\ \bibinfo {author} {\bibfnamefont
  {Y.}~\bibnamefont {Jia}},\ }\bibfield  {title} {\bibinfo {title} {Phonon and
  thermal expansion properties in weyl semimetals mx (m= nb, ta; x= p, as): ab
  initio studies},\ }\href@noop {} {\bibfield  {journal} {\bibinfo  {journal}
  {Physical Chemistry Chemical Physics}\ }\textbf {\bibinfo {volume} {18}},\
  \bibinfo {pages} {14503} (\bibinfo {year} {2016})}\BibitemShut {NoStop}%
\bibitem [{\citenamefont {Lee}\ \emph {et~al.}(2015)\citenamefont {Lee},
  \citenamefont {Xu}, \citenamefont {Huang}, \citenamefont {Sanchez},
  \citenamefont {Belopolski}, \citenamefont {Chang}, \citenamefont {Bian},
  \citenamefont {Alidoust}, \citenamefont {Zheng}, \citenamefont {Neupane}
  \emph {et~al.}}]{lee2015fermi}%
  \BibitemOpen
  \bibfield  {author} {\bibinfo {author} {\bibfnamefont {C.-C.}\ \bibnamefont
  {Lee}}, \bibinfo {author} {\bibfnamefont {S.-Y.}\ \bibnamefont {Xu}},
  \bibinfo {author} {\bibfnamefont {S.-M.}\ \bibnamefont {Huang}}, \bibinfo
  {author} {\bibfnamefont {D.~S.}\ \bibnamefont {Sanchez}}, \bibinfo {author}
  {\bibfnamefont {I.}~\bibnamefont {Belopolski}}, \bibinfo {author}
  {\bibfnamefont {G.}~\bibnamefont {Chang}}, \bibinfo {author} {\bibfnamefont
  {G.}~\bibnamefont {Bian}}, \bibinfo {author} {\bibfnamefont {N.}~\bibnamefont
  {Alidoust}}, \bibinfo {author} {\bibfnamefont {H.}~\bibnamefont {Zheng}},
  \bibinfo {author} {\bibfnamefont {M.}~\bibnamefont {Neupane}}, \emph
  {et~al.},\ }\bibfield  {title} {\bibinfo {title} {Fermi surface
  interconnectivity and topology in weyl fermion semimetals taas, tap, nbas,
  and nbp},\ }\href@noop {} {\bibfield  {journal} {\bibinfo  {journal}
  {Physical Review B}\ }\textbf {\bibinfo {volume} {92}},\ \bibinfo {pages}
  {235104} (\bibinfo {year} {2015})}\BibitemShut {NoStop}%
\bibitem [{\citenamefont {Setyawan}\ and\ \citenamefont
  {Curtarolo}(2010)}]{SETYAWAN2010299}%
  \BibitemOpen
  \bibfield  {author} {\bibinfo {author} {\bibfnamefont {W.}~\bibnamefont
  {Setyawan}}\ and\ \bibinfo {author} {\bibfnamefont {S.}~\bibnamefont
  {Curtarolo}},\ }\bibfield  {title} {\bibinfo {title} {High-throughput
  electronic band structure calculations: Challenges and tools},\ }\href@noop
  {} {\bibfield  {journal} {\bibinfo  {journal} {Computational Materials
  Science}\ }\textbf {\bibinfo {volume} {49}},\ \bibinfo {pages} {299}
  (\bibinfo {year} {2010})}\BibitemShut {NoStop}%
\bibitem [{\citenamefont {Giustino}(2017)}]{Giustino2017}%
  \BibitemOpen
  \bibfield  {author} {\bibinfo {author} {\bibfnamefont {F.}~\bibnamefont
  {Giustino}},\ }\bibfield  {title} {\bibinfo {title} {Electron-phonon
  interactions from first principles},\ }\href
  {https://doi.org/10.1103/RevModPhys.89.015003} {\bibfield  {journal}
  {\bibinfo  {journal} {Rev. Mod. Phys.}\ }\textbf {\bibinfo {volume} {89}},\
  \bibinfo {pages} {015003} (\bibinfo {year} {2017})}\BibitemShut {NoStop}%
\bibitem [{\citenamefont {Brunin}\ \emph {et~al.}(2020)\citenamefont {Brunin},
  \citenamefont {Miranda}, \citenamefont {Giantomassi}, \citenamefont {Royo},
  \citenamefont {Stengel}, \citenamefont {Verstraete}, \citenamefont {Gonze},
  \citenamefont {Rignanese},\ and\ \citenamefont {Hautier}}]{Brunin2020b}%
  \BibitemOpen
  \bibfield  {author} {\bibinfo {author} {\bibfnamefont {G.}~\bibnamefont
  {Brunin}}, \bibinfo {author} {\bibfnamefont {H.~P.~C.}\ \bibnamefont
  {Miranda}}, \bibinfo {author} {\bibfnamefont {M.}~\bibnamefont
  {Giantomassi}}, \bibinfo {author} {\bibfnamefont {M.}~\bibnamefont {Royo}},
  \bibinfo {author} {\bibfnamefont {M.}~\bibnamefont {Stengel}}, \bibinfo
  {author} {\bibfnamefont {M.~J.}\ \bibnamefont {Verstraete}}, \bibinfo
  {author} {\bibfnamefont {X.}~\bibnamefont {Gonze}}, \bibinfo {author}
  {\bibfnamefont {G.-M.}\ \bibnamefont {Rignanese}},\ and\ \bibinfo {author}
  {\bibfnamefont {G.}~\bibnamefont {Hautier}},\ }\bibfield  {title} {\bibinfo
  {title} {Phonon-limited electron mobility in si, gaas, and gap with exact
  treatment of dynamical quadrupoles},\ }\href
  {https://doi.org/10.1103/PhysRevB.102.094308} {\bibfield  {journal} {\bibinfo
   {journal} {Phys. Rev. B}\ }\textbf {\bibinfo {volume} {102}},\ \bibinfo
  {pages} {094308} (\bibinfo {year} {2020})}\BibitemShut {NoStop}%
\bibitem [{\citenamefont {Claes}\ \emph {et~al.}(2022)\citenamefont {Claes},
  \citenamefont {Brunin}, \citenamefont {Giantomassi}, \citenamefont
  {Rignanese},\ and\ \citenamefont {Hautier}}]{Claes2022}%
  \BibitemOpen
  \bibfield  {author} {\bibinfo {author} {\bibfnamefont {R.}~\bibnamefont
  {Claes}}, \bibinfo {author} {\bibfnamefont {G.}~\bibnamefont {Brunin}},
  \bibinfo {author} {\bibfnamefont {M.}~\bibnamefont {Giantomassi}}, \bibinfo
  {author} {\bibfnamefont {G.-M.}\ \bibnamefont {Rignanese}},\ and\ \bibinfo
  {author} {\bibfnamefont {G.}~\bibnamefont {Hautier}},\ }\bibfield  {title}
  {\bibinfo {title} {Assessing the quality of relaxation-time approximations
  with fully automated computations of phonon-limited mobilities},\ }\href
  {https://doi.org/10.1103/PhysRevB.106.094302} {\bibfield  {journal} {\bibinfo
   {journal} {Phys. Rev. B}\ }\textbf {\bibinfo {volume} {106}},\ \bibinfo
  {pages} {094302} (\bibinfo {year} {2022})}\BibitemShut {NoStop}%
\bibitem [{\citenamefont {Ponc{\'e}}\ \emph {et~al.}(2020)\citenamefont
  {Ponc{\'e}}, \citenamefont {Li}, \citenamefont {Reichardt},\ and\
  \citenamefont {Giustino}}]{ponce2020first}%
  \BibitemOpen
  \bibfield  {author} {\bibinfo {author} {\bibfnamefont {S.}~\bibnamefont
  {Ponc{\'e}}}, \bibinfo {author} {\bibfnamefont {W.}~\bibnamefont {Li}},
  \bibinfo {author} {\bibfnamefont {S.}~\bibnamefont {Reichardt}},\ and\
  \bibinfo {author} {\bibfnamefont {F.}~\bibnamefont {Giustino}},\ }\bibfield
  {title} {\bibinfo {title} {First-principles calculations of charge carrier
  mobility and conductivity in bulk semiconductors and two-dimensional
  materials},\ }\href@noop {} {\bibfield  {journal} {\bibinfo  {journal}
  {Reports on Progress in Physics}\ }\textbf {\bibinfo {volume} {83}},\
  \bibinfo {pages} {036501} (\bibinfo {year} {2020})}\BibitemShut {NoStop}%
\bibitem [{\citenamefont {Behnia}(2015)}]{behnia2015fundamentals}%
  \BibitemOpen
  \bibfield  {author} {\bibinfo {author} {\bibfnamefont {K.}~\bibnamefont
  {Behnia}},\ }\href@noop {} {\emph {\bibinfo {title} {Fundamentals of
  thermoelectricity}}}\ (\bibinfo  {publisher} {OUP Oxford},\ \bibinfo {year}
  {2015})\BibitemShut {NoStop}%
\bibitem [{\citenamefont {Li}(2015)}]{li2015electrical}%
  \BibitemOpen
  \bibfield  {author} {\bibinfo {author} {\bibfnamefont {W.}~\bibnamefont
  {Li}},\ }\bibfield  {title} {\bibinfo {title} {Electrical transport limited
  by electron-phonon coupling from boltzmann transport equation: An ab initio
  study of si, al, and mos$_2$},\ }\href
  {https://doi.org/10.1103/physrevb.92.075405} {\bibfield  {journal} {\bibinfo
  {journal} {Physical Review B}\ }\textbf {\bibinfo {volume} {92}},\ \bibinfo
  {pages} {075405} (\bibinfo {year} {2015})}\BibitemShut {NoStop}%
\bibitem [{\citenamefont {Fiorentini}\ and\ \citenamefont
  {Bonini}(2016)}]{fiorentini2016}%
  \BibitemOpen
  \bibfield  {author} {\bibinfo {author} {\bibfnamefont {M.}~\bibnamefont
  {Fiorentini}}\ and\ \bibinfo {author} {\bibfnamefont {N.}~\bibnamefont
  {Bonini}},\ }\bibfield  {title} {\bibinfo {title} {Thermoelectric
  coefficients of $n$-doped silicon from first principles via the solution of
  the boltzmann transport equation},\ }\href
  {https://doi.org/10.1103/PhysRevB.94.085204} {\bibfield  {journal} {\bibinfo
  {journal} {Phys. Rev. B}\ }\textbf {\bibinfo {volume} {94}},\ \bibinfo
  {pages} {085204} (\bibinfo {year} {2016})}\BibitemShut {NoStop}%
\bibitem [{\citenamefont {Eiguren}\ and\ \citenamefont
  {Ambrosch-Draxl}(2008)}]{Eiguren2008}%
  \BibitemOpen
  \bibfield  {author} {\bibinfo {author} {\bibfnamefont {A.}~\bibnamefont
  {Eiguren}}\ and\ \bibinfo {author} {\bibfnamefont {C.}~\bibnamefont
  {Ambrosch-Draxl}},\ }\bibfield  {title} {\bibinfo {title} {Wannier
  interpolation scheme for phonon-induced potentials: Application to bulk
  ${\text{mgb}}_{2}$, w, and the $(1\ifmmode\times\else\texttimes\fi{}1)$
  h-covered w(110) surface},\ }\href
  {https://doi.org/10.1103/PhysRevB.78.045124} {\bibfield  {journal} {\bibinfo
  {journal} {Phys.~Rev.~B}\ }\textbf {\bibinfo {volume} {78}},\ \bibinfo
  {pages} {045124} (\bibinfo {year} {2008})}\BibitemShut {NoStop}%
\bibitem [{\citenamefont {Shankland}(1971)}]{Shankland1971}%
  \BibitemOpen
  \bibfield  {author} {\bibinfo {author} {\bibfnamefont {D.~G.}\ \bibnamefont
  {Shankland}},\ }\bibfield  {title} {\bibinfo {title} {Fourier transformation
  by smooth interpolation},\ }\href {https://doi.org/10.1002/qua.560050857}
  {\bibfield  {journal} {\bibinfo  {journal} {International Journal of Quantum
  Chemistry}\ }\textbf {\bibinfo {volume} {5}},\ \bibinfo {pages} {497}
  (\bibinfo {year} {1971})}\BibitemShut {NoStop}%
\bibitem [{\citenamefont {Koelling}\ and\ \citenamefont
  {Wood}(1986)}]{Koelling1986}%
  \BibitemOpen
  \bibfield  {author} {\bibinfo {author} {\bibfnamefont {D.}~\bibnamefont
  {Koelling}}\ and\ \bibinfo {author} {\bibfnamefont {J.}~\bibnamefont
  {Wood}},\ }\bibfield  {title} {\bibinfo {title} {On the interpolation of
  eigenvalues and a resultant integration scheme},\ }\href
  {https://doi.org/10.1016/0021-9991(86)90261-5} {\bibfield  {journal}
  {\bibinfo  {journal} {Journal of Computational Physics}\ }\textbf {\bibinfo
  {volume} {67}},\ \bibinfo {pages} {253} (\bibinfo {year} {1986})}\BibitemShut
  {NoStop}%
\bibitem [{\citenamefont {Madsen}\ and\ \citenamefont
  {Singh}(2006)}]{Madsen2006}%
  \BibitemOpen
  \bibfield  {author} {\bibinfo {author} {\bibfnamefont {G.~K.}\ \bibnamefont
  {Madsen}}\ and\ \bibinfo {author} {\bibfnamefont {D.~J.}\ \bibnamefont
  {Singh}},\ }\bibfield  {title} {\bibinfo {title} {{BoltzTraP.} a code for
  calculating band-structure dependent quantities},\ }\href
  {https://doi.org/10.1016/j.cpc.2006.03.007} {\bibfield  {journal} {\bibinfo
  {journal} {Computer Physics Communications}\ }\textbf {\bibinfo {volume}
  {175}},\ \bibinfo {pages} {67} (\bibinfo {year} {2006})}\BibitemShut
  {NoStop}%
\bibitem [{\citenamefont {Madsen}\ \emph
  {et~al.}(2018{\natexlab{a}})\citenamefont {Madsen}, \citenamefont {Carrete},\
  and\ \citenamefont {Verstraete}}]{Madsen2018}%
  \BibitemOpen
  \bibfield  {author} {\bibinfo {author} {\bibfnamefont {G.~K.~H.}\
  \bibnamefont {Madsen}}, \bibinfo {author} {\bibfnamefont {J.}~\bibnamefont
  {Carrete}},\ and\ \bibinfo {author} {\bibfnamefont {M.~J.}\ \bibnamefont
  {Verstraete}},\ }\bibfield  {title} {\bibinfo {title} {{BoltzTraP2, a program
  for interpolating band structures and calculating semi-classical transport
  coefficients}},\ }\href {https://doi.org/10.1016/j.cpc.2018.05.010}
  {\bibfield  {journal} {\bibinfo  {journal} {Comput. Phys. Commun.}\ }\textbf
  {\bibinfo {volume} {231}},\ \bibinfo {pages} {140} (\bibinfo {year}
  {2018}{\natexlab{a}})}\BibitemShut {NoStop}%
\bibitem [{\citenamefont {Madsen}\ \emph
  {et~al.}(2018{\natexlab{b}})\citenamefont {Madsen}, \citenamefont {Carrete},\
  and\ \citenamefont {Verstraete}}]{BoltzTraP2}%
  \BibitemOpen
  \bibfield  {author} {\bibinfo {author} {\bibfnamefont {G.~K.~H.}\
  \bibnamefont {Madsen}}, \bibinfo {author} {\bibfnamefont {J.}~\bibnamefont
  {Carrete}},\ and\ \bibinfo {author} {\bibfnamefont {M.~J.}\ \bibnamefont
  {Verstraete}},\ }\bibfield  {title} {\bibinfo {title} {{BoltzTraP2}, a
  program for interpolating band structures and calculating semi-classical
  transport coefficients},\ }\href {https://doi.org/10.1016/j.cpc.2018.05.010}
  {\bibfield  {journal} {\bibinfo  {journal} {Comput. Phys. Commun.}\ }\textbf
  {\bibinfo {volume} {231}},\ \bibinfo {pages} {140 } (\bibinfo {year}
  {2018}{\natexlab{b}})}\BibitemShut {NoStop}%
\end{thebibliography}%
